\documentclass[aps,reprint,twocolumn,nofootinbib,nobibnotes, amsmath,amssymb, prd]{revtex4-2}

\usepackage[utf8]{inputenc} % Allows direct input of ñ, á, é, etc.
\usepackage[T1]{fontenc}  % Correct accent rendering in output
\usepackage{lmodern}
\usepackage{graphicx}
\usepackage{amsmath,amssymb}
\usepackage{hyperref}

\usepackage{color}
\usepackage{physics}
\usepackage{enumerate}

\hypersetup{
    colorlinks=true,
    linkcolor=blue,
    citecolor=blue,
    urlcolor=blue,
    pdftitle={Your Review Title},
    pdfauthor={Your Name}
}

\usepackage{multirow}
\usepackage[normalem]{ulem}
\useunder{\uline}{\ul}{}
\usepackage{float}
\usepackage{dsfont}
\usepackage{fancyhdr}
\usepackage{lipsum}
\usepackage{cancel}
\usepackage{array}
\usepackage[utf8]{inputenc}
\usepackage{tocloft}

\usepackage{tikz}
\usetikzlibrary{patterns}
\usepackage[compat=1.1.0]{tikz-feynman}

\tikzfeynmanset{compat=1.1.0, inline=(v.base)} 

\usepackage{silence}
\usepackage{esvect} % Para flechas de vector
\usepackage{braket} % Para notación bra-ket
\usepackage{slashed} %Para  notacion slash
\usepackage{feynmp-auto} %Para diag de Feynm

\usepackage{color}

\definecolor{mygreen}{RGB}{0,128,0}
\definecolor{mybrown}{RGB}{153,102,51}

\begin{document}

\hfill{\bf IPARCOS-UCM-25-057}
\vspace*{0.5cm}
\title{A Consistent Path Integral Approach to Higher Derivative Oscillators}    
\vspace*{0.2cm}
\author{Jose A. R. Cembranos, Eric G. Hemon, Juan J. Sanz-Cillero}
\affiliation{Departamento de F\'\i sica Te\'orica and Instituto de F\'\i sica de Part\'\i culas y del Cosmos (IPARCOS),
\\ 
Universidad Complutense de Madrid,
Plaza de las Ciencias 1, 28040-Madrid, Spain}

\begin{abstract}
\vspace*{0.2cm}
In this work, we study the Quantum Field Theory version of the higher derivative Pais-Uhlenbeck oscillator. We quantize canonically this system and construct its Fock space, as well as study its path integral.
We demonstrate that the inclusion of canonical coordinates in the path integral necessarily introduces a new field—a Lagrange multiplier— which is essential for the consistent application of these coordinates in the canonical quantization framework.
Finally, we analyze the improved ultraviolet convergence of the Green functions that this theory exhibits in the presence of an interaction.
\\ \vspace*{0.5cm} \phantom{GHOST}
\end{abstract}

\maketitle

%\tableofcontents

\pagestyle{plain}

\section{Introduction and motivation}

Quantum Field Theories (QFT) are nowadays an indispensable tool for modern physics. As a case in point, the Standard Model of particles, the most predictive theory in the history of physics, is formulated within this framework. It is thus essential to achieve a proper understanding of all types of fields that can be consistently formulated within a QFT.

Among all the fields one can construct in a QFT, we will focus upon a particular type: ghosts. These fields represent particles with negative energies (or equivalently, as we will see, negative squared norms). One might think unphysical particles like these cannot be important, but it suffices to see their central role in the Faddeev-Popov method \cite{Faddeev:2010zz},
which allows the path integral description of non-Abelian gauge theories such as Quantum Chromodynamics, to see that this is not true. In our case, we will focus on
ghosts originated by
higher-derivative theories.

These scenarios are going to include higher-than-usual time derivative terms in the Lagrangian. In the context of QFT, this typically means a field $\phi$ with a kinetic term that includes a component $\phi\Box^2\phi$ or higher powers of the d'Alembert operator. The price to pay for having this type of kinetic term is the appearance of the aforementioned ghosts, which leads to instabilities in the system.
However, several recent theoretical models are of this kind (see, e.g., \cite{Salvio:2014soa}), motivating ongoing work to render these Quantum Field Theories consistent.

One of such examples is quadratic gravity, a modified gravity theory. The Hilbert-Einstein action, which describes General Relativity as a Quantum Field Theory, is not renormalizable \cite{Buchbinder}, which means it cannot describe quantum gravity at all energy scales. The most direct way to solve this problem is by modifying the Hilbert-Einstein action by adding all possible terms quadratic in the scalar curvature: the square of the Riemann, Ricci and scalar curvature (sometimes a term $\Box R$ is also included, where $R$ is the scalar curvature). This precisely yields the quadratic gravity action. The higher-derivative terms would generate small corrections to General Relativity below the Planck mass (that is, below the cut-off of the Hilbert-Einstein action), while above this scale they would be of high importance.
The renormalizability of this theory is due to the fact that higher derivative kinetic terms induce a more convergent ultraviolet (UV) behaviour of the propagator ~\cite{Stelle:1976gc, Stelle:1977ry}. However, as said before, quadratic gravity has the same problem as any higher-derivative theory: the appearance of ghost modes~\cite{Salvio:2018crh}.

The objective of this work is to lay a groundwork for the study of higher-derivative theories. We will focus on the QFT version of the Pais-Uhlenbeck oscillator \cite{Gibbons:2019lmj}. We will thoroughly study how to quantize this system in canonical formalism and its equivalent path integral description,
as well as examine how it may behave when mediating an interaction.

This paper is structured as follows. In Sec.~\ref{seccion2}, we study the Pais-Uhlenbeck oscillator in classical mechanics and its QFT counterpart, implementing canonical quantization in both Ostrogradsky \cite{Chen:2012au} and Hawking–Hertog \cite{Hawking:2001yt} coordinates. Sec.~\ref{seccion3} analyzes the path integral generating functional of this QFT. We show that introducing a Lagrange multiplier is necessary for the consistent use of canonical coordinates, and demonstrate its full equivalence with the canonical formalism quantization analysis in the previous Section.
Finally, in Sec.~\ref{cosas_perturbativas}, we incorporate a non-derivative interaction to our free Pais–Uhlenbeck Quantum Field Theory. We use path integral to show the equivalence between the theory in its canonical-coordinate formulation and its counterpart in terms of a single field $\phi$ with a higher-derivative kinetic term. We further show that this general
type of interactions are renormalizable and provide several illustrative examples. Our final conclusions are provided in Sec.~\ref{sec:conclusions}. Some technical details have been relegated to the Appendices.

\section{Higher-derivative Lagrangians: Emergence of ghost modes}
\label{seccion2}

\subsection{Ghosts in classical, non-relativistic theories}
\label{sec:ClassMec}

The non-quantized and non-relativistic classical mechanics version of the Lagrangian we will work on is:
\begin{equation}
    L=-\frac{a}{2}\Ddot{x}^2+\frac{b}{2}\Dot{x}^2-\frac{c}{2}x^2,
    \label{PU_classical}
\end{equation}
where $a, b, c\in \mathbb{R}$, $a\neq0$ and $x(t)\in\mathbb{R}$. This system is called Pais-Uhlenbeck oscillator \cite{Pais:1950za, Pavsic:2016ykq,Pavsic:2013noa, Smilga:2008pr}. Note the dependence of the Lagrangian is $L(x,\dot{x},\Ddot{x})$, having the non-standard dependence $\Ddot{x}$, hence being a higher
-derivative theory.

The kinematic equation is easily derived: $a\!\!\overset{\ ....}{x}+b\Ddot{x}+cx=0$. We can solve it using the characteristic polynomial technique. The nature of the solution changes depending on the type of roots we are dealing with. For example, let us suppose $b^2-4ac>0$. Then roots will lie in either $\mathbb{R}$ or $i\mathbb{R}$. In the first case, we will have hyperbolic solutions, which correspond to tachyonic modes, which are not of our interest. The second solutions are trigonometric, and will be the main focus of this article. The solutions for $b^2-4ab<0$, as well as the limit cases $b^2-4ac=0$, are even more exotic: they contain mixed cases (with both hyperbolic and trigonometric functions) and polynomial factors, respectively.

We will analyze the solution of the Pais-Uhlenbeck oscillator with values that satisfy:
\begin{equation}
    b^2-4ac>0\ ,\ \  \omega_{1,2}^2\equiv  \frac{b\pm\sqrt{b^2-4ac}}{2a}\geq 0\,.
    \label{condicion_positividad}
\end{equation}
Note that these conditions enforce $\omega^2_{1,2}$ to be real, distinct\footnote{The equal-frequency case will not be studied in this article, since analyzing this system as the limit $\Delta\omega^2\to0^+$ of the unequal-frequency case proves to be a rather complex task \cite{Mannheim:2004qz, Smilga:2005gb}.} and non-negative. Although they do not determine the sign of $\Delta\omega^2 \equiv \omega_1^2-\omega_2^2$, they ensure that $a\Delta\omega^2>0$.
Nonetheless, in many applications we will see that $a$ is often taken to be positive, producing the hierarchy $\omega_1^2>\omega_2^2\geq0$.

By means of these frequencies $\omega_{1,2}$ and up to a total time derivative term, it is possible to rewrite our Lagrangian~(\ref{PU_classical}) in the equivalent form:
\begin{equation}
    L=-\frac{a}{2}x(t)\left(\frac{d^2}{dt^2}+\omega_1^2\right)\left(\frac{d^2}{dt^2}+\omega_2^2\right)x(t)\, .
    \label{PU_classical_mejornotacion}
\end{equation}

The general solution to the equations of motion (EoM) is given by:
\begin{equation}
x(t) = A_1\cos(\omega_1 t+ \delta_1)    + A_2\cos(\omega_2 t+\delta_2)  \, .
\end{equation}

To better understand the nature of the type 1 and 2 modes of the system, we must study their contributions to the Hamiltonian. In order to obtain this object, we will use Ostrogradsky canonical coordinates \cite{Chen:2012au}:
\begin{equation}
    x_1=x\ ,\ \ \ x_2=\Dot{x},
\label{ostrogradsky_clasico}
\end{equation}
\begin{equation}
    p_1=\frac{\partial L}{\partial \Dot{x}}-\frac{d}{dt}\frac{\partial L}{\partial \ddot{x}}=b\Dot{x}+a\overset{\ ...}{x}\ ,\ \ \ p_2=\frac{\partial L}{\partial \ddot{x}}=-a\ddot{x}\nonumber.
\end{equation}
Note that we choose $x_1=x$ and $t$ to have length dimensions (in natural units). This canonical choice corresponds to $a$ being a length and to $b$ and $c$ being inverse length and the cube of an inverse length respectively (or equivalently, $\omega_{1,2}$ being energies). This implies that $x_2$ is dimensionless and the associate canonical momenta also have their dimensionality fixed: $p_1$ has inverse-length dimensions and $p_2$ is dimensionless.

Performing a Legendre transform on the Lagrangian~(\ref{PU_classical}) yields the Hamiltonian:
\begin{align}
H(x_1,x_2,p_1,p_2)=& p_1\dot{x}_1+p_2\dot{x}_2-L\label{hamiltoniano_clasico}\\
    =&-\frac{1}{2a}p_2^2-\frac{b}{2}x_2^2+\frac{c}{2}x_1^2+p_1x_2.\notag
\end{align}
Substituting the relations~(\ref{ostrogradsky_clasico}) we obtain:
{\small\begin{equation}
H(x,\dot{x},\Ddot{x},\dddot{x})=a\left(\dddot{x} \dot{x}-\frac{1}{2}\Ddot{x}^2+\frac{\omega_1^2+\omega_2^2}{2}\dot{x}^2+\frac{\omega_1^2\omega_2^2}{2}x^2\right).
\end{equation}}

\noindent Evaluating this Hamiltonian in the solutions $x(t)$ from the equations of motion, we finally get:
\begin{equation}
H=   \frac{a\Delta\omega^2}{2}\, \left[\omega_2^2 A_2^2 -\omega_1^2 A_1^2\right] \, .
\end{equation}
Observe the Hamiltonian is time-independent. We can see the type-2 mode contributes positively to energy, whereas type-1 contributes negatively. For this reason, type-1 mode is said to be a ghost mode. Noting that the amplitudes of the modes can be arbitrarily large, we can then conclude the Hamiltonian is not bounded from above or from below. This is an example of the so-called Ostrogradsky instability \cite{Ostrogradsky:1850fid}: Lagrangians with a dependence on second or higher order time derivatives have associated unbounded Hamiltonians. This is the main problem with systems containing ghosts: they lead to instabilities by extending the Hamiltonian spectrum to arbitrarily negative energies.

In the next Subsections, we will extend this analysis and  examine the Pais-Uhlenbeck oscillator within the framework of a QFT.

\subsection{Pais-Uhlenbeck Quantum Field Theory: Lagrangian density}

The Field Theory version of the Pais-Uhlenbeck oscillator is given by the Lagrangian density for a real scalar field $\phi(t,\vec{x})$:
{\small
\begin{equation}
    \mathcal{L}=-\frac{1}{2}\phi(a\Box^2+b\Box+c)\phi=-\frac{a}{2}\phi(\Box+m_1^2)(\Box+m_2^2)\phi,
    \label{lagrangiano}
\end{equation}}

\noindent where $a, m_1, m_2\in \mathbb{R}$ (the masses are obtained in a similar manner to that of the frequencies in Sec.~\ref{sec:ClassMec}) and $a\neq0$. The metric signature is taken to be $+---$.

In order to see the similarity with the classical Pais-Uhlenbeck oscillator in a more explicit form, it is convenient to express $\phi(t,\vec{x})$ in terms of its Fourier transform $\widetilde{\phi}(t,\vec{p}\hspace{0.1em})$:
\begin{align}
    \mathcal{L} = \int \frac{d^3p}{(2\pi)^3}\frac{d^3q}{(2\pi)^3}\frac{-a}{2}
    &\, \Tilde{\phi}(t,\Vec{q}\hspace{0.1em})(\partial_t^2+E_{p_1}^2) \label{casi_lagrangiano_campos} \\
    & (\partial_t^2+E_{p_2}^2)\,\Tilde{\phi}(t,\Vec{p}\hspace{0.1em})\,
    e^{-i(\Vec{p}+\Vec{q}\hspace{0.1em})\cdot\Vec{x}}, \notag
\end{align}
where we define the energies $E_{p_\alpha}\equiv\sqrt{m_\alpha^2+\Vec{p}^{\: 2}}$ and the transform,
\begin{equation}
    \phi(t,\Vec{x})=\int\frac{d^3p}{(2\pi)^3}\Tilde{\phi}(t,\Vec{p}\hspace{0.1em} )e^{-i\Vec{p}\dotproduct\Vec{x}}.
    \label{Fourier_3D}
\end{equation}

By integrating the Lagrangian density~(\ref{casi_lagrangiano_campos}) over the space coordinate $\vec{x}$, the previous expression yields the Lagrangian:
\begin{equation}
    L=\int\frac{d^3p}{(2\pi)^3}\frac{-a}{2}\Tilde{\phi}^*(t,\Vec{p}\hspace{0.1em})(\partial_t^2+E_{p_1}^2)(\partial_t^2+E_{p_2}^2)\Tilde{\phi}(t,\Vec{p}\hspace{0.1em})\, ,
    \label{lagrangiano_campos}
\end{equation}
where we took into account that $\Tilde{\phi}(t,-\Vec{p}\hspace{0.1em})=\Tilde{\phi}^*(t,\Vec{p}\hspace{0.1em})$ for a real field $\phi(t,\vec{x})$.

We can see (\ref{lagrangiano_campos}) is the complex version of a continuous set of Pais-Uhlenbeck oscillators $\Tilde{\phi}(t,\Vec{p}\hspace{0.1em})$. Indeed, decomposing $\Tilde{\phi}$ in real and imaginary parts (and integrating by parts, assuming the boundary term vanishes) we find:
{\small\begin{align}
    L &=\int\frac{d^3p}{(2\pi)^3}\frac{-a}{2}\Re\Tilde{\phi}(t,\Vec{p}\hspace{0.1em})(\partial_t^2+E_{p_1}^2)(\partial_t^2+E_{p_2}^2)\Re\Tilde{\phi}(t,\Vec{p}\hspace{0.1em})\notag\\
    &+\int\frac{d^3p}{(2\pi)^3}\frac{-a}{2}\Im\Tilde{\phi}(t,\Vec{p}\hspace{0.1em})(\partial_t^2+E_{p_1}^2)(\partial_t^2+E_{p_2}^2)\Im\Tilde{\phi}(t,\Vec{p}\hspace{0.1em}),\label{infinitos_PU}
\end{align}}

\noindent which is, comparing with (\ref{PU_classical_mejornotacion}), the sum of a continuous set of Pais-Uhlenbeck oscillators $\Re\Tilde{\phi}(t,\Vec{p}\hspace{0.1em})$ and $\Im\Tilde{\phi}(t,\Vec{p}\hspace{0.1em})$. As a final note, observe that for $\phi$ to have the usual canonical dimensions one must choose $a$ to have inverse square mass dimensions.

In conclusion, the provided Lagrangian density represents a continuous set of free Pais-Uhlenbeck oscillators (i.e., each oscillator is decoupled from all the others).
Each frequency will correspond to  the energy of a particle associated to a field with two mass modes $m_1$ and $m_2$.

\subsection{Equations of motion and propagator}
\label{subsecIIC}

Consider the previous Lagrangian density~(\ref{lagrangiano}) for  a real scalar field $\phi$, where the masses were given by the relation $m_{1,2}^2=\frac{b\pm\sqrt{b^2-4ac}}{2a}$.
In later calculations, it will be useful to keep in mind their relation with the original $a,b,c$ parameters of the Lagrangian~(\ref{lagrangiano}):
\begin{equation}
\begin{cases}
    m_1^2+m_2^2 = \dfrac{b}{a} \\[6pt]
    m_1^2 m_2^2 = \dfrac{c}{a} \\[6pt]
    a m_\alpha^4 - b m_\alpha^2 + c = 0
\end{cases}.
\label{relacionesmasas}
\end{equation}

Let us note that this is the most general free Lorentz-invariant Lagrangian for a scalar field up to order $\Box^2$.

We consider $a\neq0$ to avoid trivial cases. We will choose, too: $ab>0$ and $ac\geq0$, which are necessary and sufficient conditions for $m_1^2\neq m_2^2$ and $m_1^2,m_2^2\geq0$.\footnote{If $m_1^2= m_2^2$ we have a double pole in the propagator (extremely pathological). If $m_1^2$ and/or $m_2^2<0$ we have tachyonic modes, which we will not consider. The non-negativity and distinctness of $m_{1,2}^2$ is equivalent to  (\ref{condicion_positividad}). In fact, note that in this system $m_i^2$ plays the role of $\omega_i^2$.
} The equation of motion can be immediately read from the Lagrangian:
\begin{equation}
    a(\Box+m_1^2)(\Box+m_2^2)\phi(\textbf{x})=0.
\end{equation}

We can obtain the Green function associated with this kinematic equation by solving,
\begin{equation}
a(\Box+m_1^2)(\Box+m_2^2)G(\textbf{x})=-i\delta^{(4)}(\textbf{x}) \, .
\end{equation}
By means of a 4-dimensional Fourier transform, we obtain:
\begin{align}
    \Tilde{G}(\textbf{p})&=\frac{-i/a}{(\textbf{p}^2-m_1^2)(\textbf{p}^2-m_2^2)}
    \label{propagador}
    \\
    &=\frac{1}{a\Delta m^2}\left(\frac{i}{\textbf{p}^2-m_2^2}+\frac{-i}{\textbf{p}^2-m_1^2}\right),\nonumber
\end{align}
where we denote $\Delta m^2\equiv m_1^2-m_2^2$. Note that although we will often consider the case $\Delta m^2>0$, in principle the sign of $\Delta m^2$ is free. Nevertheless, the combination $a\Delta m^2$ will be always positive.

Observe that the propagator has two simple poles with opposite sign residues. One might
suspect that each one of them is associated with a certain type of particle. Alongside the Feynman prescription, this function is the $\phi$ propagator.

Finally, we will solve the EoM by means of a Fourier transform. Performing this transformation on both sides and imposing the reality of $\phi$, we find,
{\small\begin{align}
    \phi(\textbf{x})=\sum\limits_{\alpha=1,2}\int\frac{d^3p}{(2\pi)^3}\frac{1}{\sqrt{2E_{p_\alpha}}}\big(a_p^{\scriptscriptstyle{(\alpha)}}e^{-i\textbf{p}\dotproduct\textbf{x}}+a_p^{\scriptscriptstyle{(\alpha)} \dagger}e^{i\textbf{p}\dotproduct\textbf{x}}\big)\, ,
    \label{campophi}
\end{align}}

\noindent where we are implicitly considering that each one of the $\alpha=1,2$ terms is evaluated in $p^0=E_{p_\alpha}$ respectively. This shortened field notation will be maintained for the rest of the article. We observe that our field $\phi$ has two types of modes of completely different masses and natures, as Eq.~(\ref{propagador}) suggests.

\subsection{Quantization in Ostrogradsky canonical coordinates}
\label{sec:quantization-phi1-phi2}

In order to quantize this theory, we must choose canonical coordinates. We will employ Ostrogradsky canonical coordinates, which are (\ref{ostrogradsky_clasico}), as previously shown. In this system, they take the form:
\begin{align}
    &\phi_1=\phi \ ,\ \phi_2=\partial_0\phi,
    \label{coord_Ostrogradsky}\\&\pi_1=b\partial_0\phi+a(\partial_0^3\phi-\partial_0\nabla^2\phi) \ ,\ \pi_2=-a(\partial_0^2\phi-\nabla^2\phi).\nonumber
\end{align}
In order to compute the conjugate field momenta, we have employed the equivalent Lagrangian density,
\begin{equation}
    \mathcal{L}=-\frac{a}{2}\partial_\mu\partial^\mu\phi\partial_\nu\partial^\nu\phi+\frac{b}{2}\partial_\mu\phi\partial^\mu\phi-\frac{c}{2}\phi^2 \,.
\end{equation}
Integrating by parts, it is easy to observe that this coincides with (\ref{lagrangiano}) up to an irrelevant total divergence.

The Hamiltonian density, as usual, is defined as the Legendre transform of the Lagrangian density, $\mathcal{H}=\pi_1\partial_0\phi_1+\pi_2\partial_0\phi_2-\mathcal{L}$. Eq.~(\ref{coord_Ostrogradsky}) provides the values of $\phi$, $\partial_0\phi$, $\partial_0^2\phi$ and $\partial_0^3\phi$ in terms of the Ostrogradsky canonical coordinates. Taking this into account, we find:
\begin{align}
    \mathcal{H}=&-\frac{1}{2a}\pi_2^2-\frac{b}{2}\phi_2^2+\frac{b}{2}(\nabla\phi_1)^2\notag\\
    &+\frac{c}{2}\phi_1^2+\pi_1\phi_2+\pi_2\nabla^2\phi_1.
    \label{hamiltoniano}
\end{align}

Before studying the properties of this Hamiltonian density, we should quantize the theory. To achieve this, we promote the canonical coordinates $\phi_1$, $\phi_2$, $\pi_1$, $\pi_2$ (and thus the coefficients $a_p^{\scriptscriptstyle{(1)}}$, $a_p^{\scriptscriptstyle{(2)}}$, $a_p^{\scriptscriptstyle{(1)} \dagger}$, $a_p^{\scriptscriptstyle{(2)} \dagger}$) from functions to operators. Furthermore, we must impose that they satisfy the equal time canonical commutation relations:
\begin{align}
    [\phi_\alpha(t,\Vec{x}), \phi_\beta(t,\Vec{y})]=0 \ \ &,\ \ [\pi_\alpha(t,\Vec{x}), \pi_\beta(t,\Vec{y})]=0,\nonumber\\
    [\phi_\alpha(t,\Vec{x}), \pi_\beta(t,\Vec{y})]&=i\delta^{(3)}(\Vec{x}-\Vec{y})\delta_{\alpha\beta} \ \ \forall t.\label{relaciones_conmutacion_canonicas}
\end{align}

Employing the Fourier decomposition of the $\phi$ field from  Eq.~(\ref{campophi}) in the Ostrogradsky canonical coordinates (\ref{coord_Ostrogradsky}), we are able to write them down in terms of these Fourier coefficients $a_p^{\scriptscriptstyle{(1)}}$, $a_p^{\scriptscriptstyle{(2)}}$, $a_p^{\scriptscriptstyle{(1)} \dagger}$, $a_p^{\scriptscriptstyle{(2)} \dagger}$ (which are now operators):
{\small\begin{align}
\phi_1(\textbf{x}) &= \sum\limits_{\alpha=1,2}\int\frac{d^3p}{(2\pi)^3} \frac{1}{\sqrt{2E_{p_\alpha}}} \left(
    a_p^{\scriptscriptstyle{(\alpha)}}e^{-i\textbf{p}\dotproduct\textbf{x}}
    + a_p^{\scriptscriptstyle{(\alpha)}\dagger}e^{i\textbf{p}\dotproduct\textbf{x}}
\right),\notag \\
\phi_2(\textbf{x}) &= \sum\limits_{\alpha=1,2} \int \frac{d^3p}{(2\pi)^3} (-i) \sqrt{\frac{E_{p_\alpha}}{2}} \Big(
    a_p^{\scriptscriptstyle{(\alpha)}} e^{-i\textbf{p}\dotproduct\textbf{x}}- a_p^{\scriptscriptstyle{(\alpha)\dagger}} e^{i\textbf{p}\dotproduct\textbf{x}}
\Big),\notag \\
\pi_1(\textbf{x})  &= \sum\limits_{\alpha=1,2}\int\frac{d^3p}{(2\pi)^3} i(am_\alpha^2 - b) \sqrt{\frac{E_{p_\alpha}}{2}} \Big(
    a_p^{\scriptscriptstyle{(\alpha)}} e^{-i\textbf{p}\dotproduct\textbf{x}}
    \label{eq:Ostrogradsky-coord}
    \\
    &\hspace{15.5em}- a_p^{\scriptscriptstyle{(\alpha)\dagger}} e^{i\textbf{p}\dotproduct\textbf{x}}
\Big),\notag \\
\pi_2(\textbf{x})  &= \sum\limits_{\alpha=1,2}\int\frac{d^3p}{(2\pi)^3} \frac{am_\alpha^2}{\sqrt{2E_{p_\alpha}}} \left(
    a_p^{\scriptscriptstyle{(\alpha)}}e^{-i\textbf{p}\dotproduct\textbf{x}}
    + a_p^{\scriptscriptstyle{(\alpha)}\dagger}e^{i\textbf{p}\dotproduct\textbf{x}}
\right).\notag
\end{align}}

\noindent Substituting these expressions in the canonical commutation relations~(\ref{relaciones_conmutacion_canonicas}) we are now able to extract the corresponding commutation relations satisfied by the Fourier coefficients $a_p^{\scriptscriptstyle{(1)}}$, $a_p^{\scriptscriptstyle{(2)}}$, $a_p^{\scriptscriptstyle{(1)} \dagger}$, $a_p^{\scriptscriptstyle{(2)} \dagger}$. These commutation relations are unique, for there are as many independent commutators for the Ostrogradsky coordinates as there are for the Fourier coefficients, thereby making (\ref{relaciones_conmutacion_canonicas}) a determinate system of linear equations for them. We compute:\footnote{We recommend to perform the quantization at $t=0$ without any loss of generality (as (\ref{relaciones_conmutacion_canonicas}) is true, in particular, for $t=0$).}
\begin{align}
&[a_p^{\scriptscriptstyle{(\alpha)}}, a_q^{\scriptscriptstyle{(\beta)}}]=0 \ \ , \ \ [a_p^{\scriptscriptstyle{(\alpha)}\dagger}, a_q^{\scriptscriptstyle{(\beta)}\dagger}]=0,\notag\\
&[a_p^{\scriptscriptstyle{(\alpha)}}, a_q^{\scriptscriptstyle{(\beta)}\dagger}]=0 \ \text{ for $\alpha\neq\beta$},
\label{relaciones_conmutacion_cd}\\
&[a_p^{\scriptscriptstyle{(2)}}, a_q^{\scriptscriptstyle{(2)}\dagger}]=-[a_p^{\scriptscriptstyle{(1)}}, a_q^{\scriptscriptstyle{(1)}\dagger}]=\frac{1}{a\Delta m^2}(2\pi)^3\delta^{(3)}(\Vec{p}-\Vec{q}).\nonumber
\end{align}
We remind that the signs of the non-zero commutators are unambiguously determined, since $a\Delta m^2>0$. Thus, up to positive normalization constants, type-2 particles satisfy the standard commutation relations, while type-1 particles satisfy the same relations but with an overall sign inversion.

\subsection{Hamiltonian}

To clarify the physical meaning of the type 1 and 2 modes, we must first obtain the Hamiltonian of the system in terms of the Fourier coefficients (which, as we will see in Sec.~\ref{estados_energia_biendefinida}, are essentially creation and annihilation operators).
After substituting the previously obtained expressions~(\ref{eq:Ostrogradsky-coord}) for $\phi_1$, $\phi_2$, $\pi_1$, $\pi_2$ in the Hamiltonian density~(\ref{hamiltoniano}), and integrating in position $\Vec{x}$,  we obtain,
\begin{align}
    H = \sum\limits_{\alpha,\beta}\int \frac{d^3p}{(2\pi)^3}& \Big[ A^{\scriptscriptstyle{(+)}}_{\alpha\beta}\left(a_p^{\scriptscriptstyle (\alpha)} a_{-p}^{\scriptscriptstyle (\beta)} + a_{-p}^{\scriptscriptstyle (\alpha)\dagger} a_p^{\scriptscriptstyle (\beta)\dagger} \right)\notag \\
    &+ A^{\scriptscriptstyle{(-)}}_{\alpha\beta} \left( a_p^{\scriptscriptstyle (\alpha)} a_p^{\scriptscriptstyle (\beta)\dagger} + a_p^{\scriptscriptstyle (\alpha)\dagger} a_p^{\scriptscriptstyle (\beta)} \right) \Big]\, ,
\end{align}
where,
{\small\begin{align}
A^{\scriptscriptstyle{(\pm)}}_{\alpha\beta} &= \frac{1}{2\sqrt{E_{p_\alpha}E_{p_\beta}}} \left( \frac{1}{2} \left(c - am_\alpha^2 m_\beta^2 \right) \right. \\
&\quad \left. \pm \left(am_\alpha^2 - \frac{b}{2}\right) \left(E_{p_\alpha} E_{p_\beta} \mp \Vec{p}^{\:2} \right) \right) \,.\notag
\end{align}}

\noindent We have taken $t=0$ without loss of generality, as the action is time-translation independent (i.e., $\partial_t H=0$).

It is easy to verify using (\ref{relaciones_conmutacion_cd}) that the antisymmetric part of $A^{\scriptscriptstyle{(\pm)}}_{\alpha\beta}$ does not contribute to the Hamiltonian. After extracting the symmetric part of these objects and taking into account (\ref{relacionesmasas}), the Hamiltonian of the system is finally obtained:
\begin{align}
    H&=a\Delta m^2\int\frac{d^3p}{(2\pi)^3}\frac{E_{p_2}}{2}(a_p^{\scriptscriptstyle(2)}a_{p}^{\scriptscriptstyle(2)\dagger}+a_{p}^{\scriptscriptstyle(2)\dagger}a_{p}^{\scriptscriptstyle(2)})\notag\\
    &-a\Delta m^2\int\frac{d^3p}{(2\pi)^3}\frac{E_{p_1}}{2}(a_p^{\scriptscriptstyle(1)}a_{p}^{\scriptscriptstyle(1)\dagger}+a_{p}^{\scriptscriptstyle(1)\dagger}a_{p}^{\scriptscriptstyle(1)}).
    \label{hamiltoniano_final}
\end{align}

Now it is clear the true nature of the type 1 and 2 modes, which are decoupled: they are both Klein-Gordon modes with different masses $m_1$ and $m_2$. As we will later point out when we define the one-particle states associated to these modes in Sec.~\ref{estados_energia_biendefinida}, one of them is a standard mode and the other one is a ghost mode. The latter is precisely the consequence of introducing higher-derivative terms in the Lagrangian density of this scalar field theory.

Note that our Fourier-mode coordinates have been good enough in the sense that the Hamiltonian is already in its diagonal form, that is, without any other crossed term of the form $a^{\scriptscriptstyle(1)}a^{\scriptscriptstyle(2)}$, $a^{\scriptscriptstyle(1)}a^{\scriptscriptstyle(2)\dagger}$ or similar. If this separation had not happened, we would have problems to interpret the vacuum excitations generated by these operators $a^{\scriptscriptstyle(\alpha)\dagger}$ as Hamiltonian eigenstates, and thus, as the true energy modes of the system (as we will do in Sec.~\ref{estados_energia_biendefinida}). In this case, the Hamiltonian should be diagonalized via a linear transformation of the Fourier-mode coordinates.

Although the Ostrogradsky canonical coordinates are perfectly valid, we observe they mix standard and ghost modes that are not coupled, as they appear independently in the Hamiltonian. It would be convenient to introduce a new set of canonical coordinates in which this fact becomes manifest. The next Section will detail how to construct these coordinates starting from the Ostrogradsky ones.

\subsection{Quantization in Hawking-Hertog canonical coordinates}
\label{sec:quantization-psi1-psi2}

Based on the derivation from Hawking and Hertog~\cite{Hawking:2001yt}, we define the coordinates:
\begin{equation}
    \psi_{1 (2)}=\sqrt{\frac{a}{\Delta m^2}}\left(\Box+m_{2(1)}^2\right)\phi.
    \label{hawking-hertog}
\end{equation}
It is easy to verify substituting these definitions in (\ref{lagrangiano}) that the Lagrangian density is equal to:
\begin{equation}
    \mathcal{L}=-\frac{1}{2}\psi_2(\Box+m_2^2)\psi_2+\frac{1}{2}\psi_1(\Box+m_1^2)\psi_1.
    \label{lagrangiano_Hawking}
\end{equation}
That is, these new fields are obtained by projecting the original $\phi$ field in its standard and ghost components. These components are a standard Klein-Gordon field ($\psi_2$) and a ghost Klein-Gordon field ($\psi_1$). The only difference between the latter and the former is the sign of its kinetic term.

We define the canonical conjugate momenta of  $\psi_1$ and $\psi_2$ ($\sigma_1$ and $\sigma_2$ respectively) in the usual way, as in theories without higher derivatives. Note that $\sigma_2=\partial_0\psi_2$ and $\sigma_1=-\partial_0\psi_1$. The non-standard minus sign of $\sigma_1$ is the origin of the strange commutation relations of the ghost mode. We can find the transformation that links the original coordinates ($\phi_1$, $\phi_2$, $\pi_1$, $\pi_2$) and these new coordinates ($\psi_1$, $\psi_2$, $\sigma_1$, $\sigma_2$). The result is:
\begin{equation}
\psi_{1(2)}=\sqrt{\frac{a}{\Delta m^2}}\left(m_{2(1)}^2\phi_1-\frac{1}{a}\pi_2\right),
\end{equation}
\begin{equation}
    \sigma_\alpha=(-1)^\alpha\sqrt{\frac{a}{\Delta m^2}}\left(\frac{1}{a}\pi_1-m_\alpha^2\phi_2\right).
\end{equation}
Moreover, using these relations and (\ref{relaciones_conmutacion_canonicas}) we have verified that these new coordinates satisfy the same commutation relations as the Ostrogradsky ones. Thus, the transformation ($\phi_1$, $\phi_2$, $\pi_1$, $\pi_2$) to ($\psi_1$, $\psi_2$, $\sigma_1$, $\sigma_2$) is canonical.

The Hamiltonian computation provides a final consistency check of the coordinate transformation. Using $\mathcal{H}=\sigma_1\partial_0\psi_1+\sigma_2\partial_0\psi_2-\mathcal{L}$, we find:
\begin{align}
    \mathcal{H}&=\frac{1}{2}\left(\sigma_2^2+(\nabla\psi_2)^2+m_2^2\psi_2^2)\right)\notag \\
    &-\frac{1}{2}\left(\sigma_1^2+(\nabla\psi_1)^2+m_1^2\psi_1^2)\right).
\end{align}
This is just the sum of two usual Klein-Gordon energies but having the type-1 term a different sign, which clearly
leads to the same outcome we found in Eq.~(\ref{hamiltoniano_final}).

\subsection{Possible extensions to even higher-derivatives}
Consider now the Lagrangian \cite{Masterov:2015ija, Masterov:2016jft}:
\begin{equation}
    \mathcal{L}=-\frac{a}{2}\phi(\Box+m_1^2)\ldots(\Box+m_n^2)\phi,
\label{derivadasmuysuperiores}
\end{equation}
where $n\in\mathbb{N}_+$, $m_1^2<\ldots<m_n^2$ and $a>0$. Doing the following change of coordinates:
{\small\begin{equation}
    \psi_i=\sqrt{\frac{a}{\prod\limits_{i\neq j=1}^n\abs{m_j^2-m_i^2}}}\left[\prod\limits_{i\neq j=1}^n(\Box+m_j^2)\right]\phi,
\end{equation}}

\noindent with $i=1,\ldots,n$; we obtain:
\begin{align}
    \mathcal{L}=&-\frac{1}{2}\psi_1(\Box+m_1^2)\psi_1+\frac{1}{2}\psi_2(\Box+m_2^2)\psi_2+\ldots\nonumber\\
    &\ldots+(-1)^n\frac{1}{2}\psi_n(\Box+m_n^2)\psi_n.
\end{align}
One consistency check for this statement is obtained by decomposing the propagator of the (\ref{derivadasmuysuperiores}) $\phi$ field, which is (solving the Green function of the kinematic equation):
\begin{equation}
    \Tilde{G}(\textbf{p})=\frac{(-1)^{n+1}\; i/a}{(p^2-m_1^2)\ldots(p^2-m_n^2)},
\end{equation}
in a sum of $n$ propagators with simple poles (using the partial fraction decomposition technique on this quotient of polynomials).

\section{Path integral and Fock space construction}
\label{seccion3}

\subsection{Inclusion of canonical coordinates in the path integral}

Let us see the consequences of introducing canonical coordinates in our higher-derivative theory from a path integral perspective. Starting with a simpler system, we quantize our classical mechanics Hamiltonian (\ref{hamiltoniano_clasico}) promoting the phase space coordinates $(x_1,x_2,p_1,p_2)$ to operators which satisfy:
\begin{equation}
    [x_i,x_j]=0\ \ \ ,\ \ \ [p_i,p_j]=0\ \ \ ,\ \ \ [x_i,p_j]=i\delta_{ij}\ .
\end{equation}
This defines the quantized Hamiltonian (\ref{hamiltoniano_clasico}) without ambiguity because $p_1x_2=x_2p_1$. Moreover, this Hamiltonian is Weyl-ordered, so if we do the usual Quantum Mechanics path integral construction,\footnote{For more details, see
, for instance, Sec.~9.1 of Peskin and Schroeder's book \cite{peskin}.} we get:
\begin{align}
&U(x,x';T)\equiv\bra{x_1',x_2'}e^{-iH(x_1,x_2,p_1,p_2)T}\ket{x_1,x_2}\\
&=\int Dx_1(t)Dx_2(t)Dp_1(t)Dp_2(t)\notag\\
&\ \ \ \ \times e^{i\int_0^Tdt(p_1\dot{x}_1+p_2\dot{x}_2-H(x_1,x_2,p_1,p_2))},\notag
\label{QM_pathintegral}
\end{align}
where inside the path integral we have functions instead of operators. Initial and final values of $x_i(0)$ and $x_i(T)$ are fixed by $x_i$ and $x_i'$ given by $\ket{x_1,x_2}$ and $\ket{x_1',x_2'}$ respectively. Strictly speaking, we should employ a rigorous treatment of the path integral via spacetime discretization; however, we will omit such details.

Now we are going to integrate out as many coordinates as possible. For that purpose, we write the exponent of the integrand as follows:
\begin{align}
&p_1\dot{x}_1+p_2\dot{x}_2-H(x_1,x_2,p_1,p_2)=\\
&p_1(\dot{x}_1- x_2)+\left(\frac{1}{\sqrt{2a}}p_2+\sqrt{\frac{a}{2}}\dot{x}_2\right)^2+ L(x_1,x_2,\dot{x}_2)\notag,
\end{align}
being $L(x,\dot{x},\ddot{x})$ Lagrangian (\ref{PU_classical}). Note the integration of $p_1$ results in a functional Dirac delta. Thus, the integration of $p_1$ also integrates out $x_2$ because, as we can see, it imposes $x_2=\dot{x}_1$. Finally, performing the change of variables $p_2'=\frac{1}{\sqrt{2a}}p_2+\sqrt{\frac{a}{2}}\dot{x}_2$, we can integrate out $p_2'$. In conclusion, the path integral can be written in the following way (ignoring irrelevant global constants):
\begin{equation}
    U(x,x';T)=\int Dx(t)e^{i\int_0^T dtL(x,\dot{x},\ddot{x})}.
    \label{path_integral_QM}
\end{equation}
As said, initial and final positions and velocities are fixed by $x(0)=x_1$, $x(T)=x'_1$ and $\dot{x}(0)=x_2$, $\dot{x}(T)=x'_2$ respectively.

A Quantum Field Theory is completely described by the generating functional. The Field Theory of the Pais-Uhlenbeck oscillator, as shown in (\ref{infinitos_PU}), is essentially the sum of infinite Pais-Uhlenbeck oscillators, two per energy mode. Thus, the path integral that characterizes this system is the same as the one we just saw, just with two functional integrals per independent Fourier mode (one for $\Re\Tilde{\phi}(t,\Vec{p}\hspace{0.1em})$ and one for $\Im\Tilde{\phi}(t,\Vec{p}\hspace{0.1em})$, both subject to the correspondent restrictions imposed by the reality of the $\phi$ field). This can be understood as an integration in $\phi(\mathbf{x})$. If, additionally, we insert a source term, we can formally define the aforementioned functional generator:
\begin{equation}
    W[J]=\int D\phi e^{i\int d^4\bold{x}\frac{-a}{2}\phi(\Box+m_1^2)(\Box+m_2^2)\phi+i\int d^4\bold{x}J\phi}.
\label{generador_funcional}
\end{equation}
Indeed, parting from the Quantum Field Theory path integral in the Ostrogradsky coordinates form as before, we reach this expression when integrating out all coordinates but $\phi_1$. The values of the field and its first temporal derivative are usually also fixed to some spatial function in the initial and final times of the path.  Note in (\ref{generador_funcional}) we have taken these initial and final times to be infinitely in the past and future, respectively.

However, we note that in the process of introducing the Hawking-Hertog canonical coordinates in the (\ref{generador_funcional}) path integral, an extra term in the Lagrangian appears \cite{Mannheim:2004qz}. Indeed, to make use of these coordinates,
we will insert the following functional identity in the generating functional:
\begin{align}
    1=\int D\psi_1 D\psi_2 &\, \delta\left(\psi_1-\sqrt{\frac{a}{\Delta m^2}}(\Box+m_2^2)\phi\right)\notag\\
    &\delta\left(\psi_2-\sqrt{\frac{a}{\Delta m^2}}(\Box+m_1^2)\phi\right)\, .
\label{eq:identity-deltas}
\end{align}
Moreover, in practice, we will conveniently rewrite the product of the two Dirac deltas by means of the relation,
\begin{equation}
   \int dx\, \delta(x-A)\delta(x-B) \, =\, \delta(A-B)\, =\, \int \frac{dz}{2\pi}\, e^{iz(A-B)}\, .
\label{eq:delta-relation}
\end{equation}
Note that, through this trick, the identity, which is first expressed as two integrations and two deltas, can be finally written as two integrations with no deltas, at the price of the appearance of an additional exponential term. Furthermore, we want to remind that the functional always carries an irrelevant normalization constant which will absorb any purely numerical coefficients such as the $\frac{1}{2\pi}$ factor in~(\ref{eq:delta-relation}).

Introducing the identity~(\ref{eq:identity-deltas}) in the integrand of Eq.~(\ref{generador_funcional}) and then performing the integration on $\phi$ employing (\ref{eq:delta-relation}) leads to the equivalent generating functional,
\begin{align}
    &\int D\psi_1D\psi_2D\lambda\; e^{
    i\int d^4\boldsymbol{x}\,  \Big(
        -\frac{1}{2}\psi_2(\Box+m_2^2)\psi_2
        + \frac{1}{2}\psi_1(\Box+m_1^2)\psi_1
    \Big)
    }  \nonumber \\
    &\times e^{ i\int d^4\boldsymbol{x}\, \lambda\left[(\Box+m_2^2)\psi_2-(\Box+m_1^2)\psi_1\right]+
    i\int d^4\boldsymbol{x}
         J\frac{\psi_2-\psi_1}{\sqrt{a\Delta m^2}}
    }\,. \label{funcional_generador_3campos}
\end{align}
Observe in exchange for having these two independent canonical coordinates $\psi_1$ and $\psi_2$, a third independent auxiliary coordinate $\lambda$ has appeared. Note that in the case of Ostrogradsky canonical coordinates, it was $\pi_1$ that played the role of a Lagrange multiplier, as it concealed a Dirac delta. Further details on functional identities have been relegated to App.~\ref{app:further-functional-details}.
We want to emphasize that this generating functional~(\ref{funcional_generador_3campos}) is fully equivalent to the original one~(\ref{generador_funcional}). Indeed, in App.~\ref{app:apendice1}, we show that one recovers the original expression (\ref{generador_funcional}) by integrating out $\lambda$ and the combination $(\psi_1+\psi_2)$.

We point out that this generating functional, which is the result of inserting canonical coordinates in Eq.~(\ref{generador_funcional}), does not take the form one would expect from the Lagrangian density~(\ref{lagrangiano_Hawking}), which would be:
\begin{align}
    \int D\psi_1D\psi_2\; &e^{i\int d^4\bold{x}\Big(-\frac{1}{2}\psi_2(\Box+m_2^2) \psi_2+\frac{1}{2}\psi_1(\Box+m_1^2)\psi_1\Big)}\nonumber\\
    &\times e^{i\int d^4\boldsymbol{x}\left(J_1\psi_1+J_2\psi_2\right)}.
    \label{funcional_gen_extendido}
\end{align}
Two differences appear. The first one is a reduction in the number of source terms: there exists only the source term $J$ for the original field $\phi$. This means a restriction on the Green functions that can be produced by functionally differentiating the generating functional. Indeed, the functional in Eq.~(\ref{funcional_generador_3campos}) cannot generate Green functions with any arbitrary combination $\alpha\psi_1(x)+\beta\psi_2(x)$ in the legs,
but only those with $\alpha=-\beta$. The second difference is the appearance of extra Dirac delta functions in the path integral measure, which means the inclusion of Lagrange multipliers in the Lagrangian density (denoted as $\lambda(\mathbf{x})$).

As a result of this, the generating functionals~(\ref{funcional_generador_3campos}) and~(\ref{funcional_gen_extendido}) lead to very different properties, in particular for the corresponding Euclidean actions $S_E$. By means of a Wick rotation we rewrite our space-time coordinates in the form $x^0=-ix_E^0$, $\Vec{x}=\Vec{x}_E$, $\Box= -\Box_E \equiv -  \sum\limits_{\mu=0}^3\frac{\partial^2}{\partial x_E^{\mu\ 2}}$.
In addition, to prove that the $S_E$ associated to the generating functional~(\ref{generador_funcional}) is bounded (more precisely, lower bounded for $a>0$), we will perform the Fourier transforms,
\begin{eqnarray}
    \phi(x_E)&=&\int\frac{d^4p_E}{(2\pi)^4}\Tilde{\phi}(p_E)e^{-ip_E\dotproduct x_E}\,,
    \nonumber\\
    J(x_E)&=&\int\frac{d^4p_E}{(2\pi)^4}\Tilde{J}(p_E)e^{-ip_E\dotproduct x_E}\,,
\end{eqnarray}
where $\Tilde{\phi}^*(p_E)=\Tilde{\phi}(-p_E)$ and $\Tilde{J}^*(p_E)=\Tilde{J}(-p_E)$ for real $\phi$ and $J$.
Thus, our generating functional~(\ref{generador_funcional}) takes the form,
\begin{eqnarray}
W[J]&=&\int D\phi \, e^{iS\,\, +\,\,  i \int d^4\bold{x}\, J(x)\phi(x)}
        \nonumber\\
    &=&\int D\Tilde{\phi} \, e^{-S_E\,\, +\,\, \int\frac{d^4p_E}{(2\pi)^4}\Tilde{J}(p_E)\Tilde{\phi}^*(p_E)} \,,
\end{eqnarray}
with the corresponding Euclidean action,
\begin{eqnarray}
S_E &=& \int\frac{d^4p_E}{(2\pi)^4}  \,   \frac{a}{2}(p_E^2+m_1^2)(p_E^2+m_2^2)|\Tilde{\phi}(p_E)|^2  \,.
\end{eqnarray}
Note that this Euclidean action is bounded (from above or below, depending on whether $a$ is negative or positive, respectively). However, if we take the naive generating functional~(\ref{funcional_gen_extendido}) one obtains the structure:
\begin{align}
W[J_1,J_2] &=\int D\psi_1D\psi_2\; \,
                e^{iS +  i \int d^4\bold{x}\, \left(J_1(x)\psi_1(x)+J_2(x)\psi_2(x)\right)}
            \nonumber\\
            &\hspace*{-1.6cm} = \int D\Tilde{\psi}_1D\Tilde{\psi}_2\; \,
                e^{-S_E\,\, +\,\, \int\frac{d^4p_E}{(2\pi)^4}\, \left( \Tilde{J}_1(p_E)\Tilde{\psi}_1^*(p_E)+\Tilde{J}_2(p_E)\Tilde{\psi}_2^*(p_E)\right)}\, ,
            \nonumber
\end{align}
with the Euclidean action,
\begin{eqnarray}
S_E &=&  \int \frac{d^4p_E}{(2\pi)^4}  \,
        \bigg(\frac{1}{2}(p_E^2+m_2^2) |\Tilde{\psi}_2(p_E)|^2
\label{no_acotado}
\\
    &&\qquad\qquad\qquad \, -\,  \frac{1}{2}(p_E^2+m_1^2) |\Tilde{\psi}_1(p_E)|^2 \bigg)  \,,
\nonumber
\end{eqnarray}
which is now unbounded.

The key to this different behaviour between the precise, (\ref{funcional_generador_3campos}), and naive,~(\ref{funcional_gen_extendido}), generating functionals
lies in the source terms.
As we show in full detail in App.~\ref{app:apendice1}, what is really happening here is that the Euclidean action of the equivalent generating functional~(\ref{funcional_generador_3campos}) is also unbounded. However, the fields that cause this can be integrated out, as they are not coupled to the source $J$ or to any field. Thus, they produce no effect on the Green functions, just adding an irrelevant constant factor to the global normalization of $W[J]$.

Note that the unbounded Euclidean action problem of (\ref{no_acotado}) is only worrisome if the ghost and standard modes are interacting non-trivially; in the free-field case (with a quadratic action) they can be decoupled in the generating functional through an appropriate choice of the source $J$. In Sec.~\ref{cosas_perturbativas}, we will show the equivalence of the theory in terms of $\phi$ (that is, written as in Eq.~(\ref{generador_funcional})) and the one in terms of the fields $\psi_1$, $\psi_2$ and $\lambda$ (written as in Eq.~(\ref{funcional_generador_3campos})) from the point of view of Feynman diagrams. This equivalence will be proven in the context of a non-trivial interaction of $\phi$ (which means ghost and standard modes also interact non-trivially) using a diagrammatic approach.

Now, we will explain why the appearance of this Lagrange multiplier does not affect negatively our previous discussion about the canonical quantization. For that purpose, we will redefine $\lambda(\mathbf{x})$ as $g\lambda(\mathbf{x})$ with $g\in\mathbb{R}-
\{0\}$. This will serve as a consistency check, as the results should be $g$-independent
(as we will later explicitly show).
Moreover, we will add in the Lagrangian density a term of the form $\lim\limits_{\epsilon\to0^+}\frac{-\epsilon}{2}\lambda(\Box+\eta(\epsilon)\lambda)$ with $\lim\limits_{\epsilon\to0^+}\epsilon\eta(\epsilon)=0$. Exchanging then limits with the space-time and functional integrals, we have another form for the generating functional, fully equivalent to those in~(\ref{generador_funcional}) and~(\ref{funcional_generador_3campos}):
{\small\begin{align}
    &\lim\limits_{\epsilon\to0^+}\int D\psi_1D\psi_2D\lambda\; e^{
    i\int d^4\boldsymbol{x} \Big(
        \frac{-1}{2}\psi_2(\Box+m_2^2)\psi_2
        + \frac{1}{2}\psi_1(\Box+m_1^2)\psi_1
    \Big)
    }  \nonumber \\
    &\times e^{ i\int d^4\boldsymbol{x}(-\frac{1}{2}\epsilon\lambda(\Box+\eta)\lambda+g\lambda\left[(\Box+m_2^2)\psi_2-(\Box+m_1^2)\psi_1\right])}\nonumber\\
    &\times e^{
    i\int d^4\boldsymbol{x}
         J\frac{\psi_2-\psi_1}{\sqrt{a\Delta m^2}}
    }\,.
\label{FuncionalGenerador_final}
\end{align}}

\noindent Note that, here and all through the article, we are assuming that the action of limits, integrals and functional derivatives can be in general exchanged and reordered.

Starting from the Lagrangian density associated to this functional~(\ref{FuncionalGenerador_final}), we show in the next Section that the canonical quantization is essentially the same as that previously shown in previous Secs.~\ref{sec:quantization-phi1-phi2} and~\ref{sec:quantization-psi1-psi2}.

\subsection{Canonical quantization with a Lagrange multiplier}

Let us treat canonically the Lagrangian associated to the generating functional~(\ref{FuncionalGenerador_final}):
\begin{align}
    &\mathcal{L}=-\frac{1}{2}\psi_2(\Box+m_2^2)\psi_2+\frac{1}{2}\psi_1(\Box+m_1^2)\psi_1
    \label{lagrangiano_conlambda}
    \\
    &-\frac{\epsilon}{2}\lambda(\Box+\eta)\lambda+g\lambda\left[(\Box+m_2^2)\psi_2-(\Box+m_1^2)\psi_1\right].
\notag
\end{align}
In this canonical approach we can start treating these fields as independent, because in (\ref{FuncionalGenerador_final}) the three of them are being integrated in every possible path as independent variables. Of course, restrictions will appear when we study the kinematic equations of motion. These equations can be written in the following way:
\begin{equation}
\begin{cases}
\psi_1: (\Box + m_1^2)(\psi_1 - g \lambda) = 0 & \\
\psi_2: (\Box + m_2^2)(\psi_2 - g \lambda) = 0 & \\
\lambda: \epsilon(\Box + \eta) \lambda = g \Big( (\Box + m_2^2) \psi_2 - (\Box + m_1^2) \psi_1 \Big) &
\end{cases}.
\end{equation}
Observe if $\epsilon=0$ we would have $\lambda=0$ and the canonical quantization discussion we already had would be left unchanged. However, leaving $\epsilon\to0^+$ serves two purposes: first, it helps us to better understand the role of this field; second, it allows us to manipulate the field in Feynman diagrams using the usual Feynman rules (since it enables us to write the $\lambda$ propagator), as we will see in Sec. \ref{aparicion_phi_diagramas}.

We can solve this system of equations by Fourier transforming them on both sides:
{\small\begin{align}
    &\psi_{\alpha}(\mathbf{x}) = (-1)^\alpha \int \frac{d^3\vec{p}}{(2\pi)^3}
    \sqrt{\frac{a\Delta m^2}{2E_{p_{\alpha}}}}
    \left(
        a_p^{\scriptscriptstyle{(\alpha)}} e^{-i\textbf{p}\dotproduct\textbf{x}}
        + a_p^{\scriptscriptstyle{(\alpha)}\dagger}  e^{i\textbf{p}\dotproduct\textbf{x}}
    \right)\notag
    \\
    &+g\int\frac{d^3\vec{p}}{(2\pi)^3}
    \sqrt{\frac{a\Delta m^2}{2E_{p_{\alpha}}}}
    \left(
        a_p^{\scriptscriptstyle{(3)}} e^{-i\textbf{p}\dotproduct\textbf{x}}
        + a_p^{\scriptscriptstyle{(3)}\dagger}  e^{i\textbf{p}\dotproduct\textbf{x}}
    \right),
\end{align}
\begin{equation}
    \lambda(\mathbf{x})=\int\frac{d^3\vec{p}}{(2\pi)^3}
    \sqrt{\frac{a\Delta m^2}{2E_{p_{   3  }}}}   %%%\alpha}}}}
    \left(
        a_p^{\scriptscriptstyle{(3)}} e^{-i\textbf{p}\dotproduct\textbf{x}}
        + a_p^{\scriptscriptstyle{(3)}\dagger}  e^{i\textbf{p}\dotproduct\textbf{x}}
    \right).
\end{equation}}The arbitrary prefactors are chosen so that the Fourier coefficients of the type 1 and 2 modes defined here match those defined in Sec.~\ref{subsecIIC}. Here we evaluate each $\alpha=1,2,3$ term in $p^0=E_{p_\alpha}$, where we are defining $E_{p_3}=\sqrt{m_3^2+\Vec{p}^{\: 2}}$ and $m_3^2\equiv\eta+\frac{g^2\Delta m^2}{\epsilon}$. The associate canonical momenta $\sigma_\alpha$ are defined as usual:
\begin{align}
    &\sigma_1=-\partial_0(\psi_1-g\lambda)=-\sqrt{a\Delta m^2}\times\\
    &i\int\frac{d^3p}{(2\pi)^3}\sqrt{\frac{E_{p_1}}{2}}\Big(a_p^{\scriptscriptstyle{(1)}}e^{-i\mathbf{p}\dotproduct\mathbf{x}}-a_p^{\scriptscriptstyle{(1)}\dagger }e^{i\mathbf{p}\dotproduct\mathbf{x}}\Big)\Big|_{p^0=E_{p_1}},\notag
\end{align}
\begin{align}
    &\sigma_2=\partial_0(\psi_2-g\lambda)=\sqrt{a\Delta m^2}\times\\
    &(-i)\int\frac{d^3p}{(2\pi)^3}\sqrt{\frac{E_{p_2}}{2}}\Big(a_p^{\scriptscriptstyle{(2)}}e^{-i\mathbf{p}\dotproduct\mathbf{x}}-a_p^{\scriptscriptstyle{(2)}\dagger }e^{i\mathbf{p}\dotproduct\mathbf{x}}\Big)\Big|_{p^0=E_{p_2}},\notag
\end{align}
\begin{align}
    &\sigma_\lambda=\epsilon\partial_0\lambda+g\partial_0(\psi_1-\psi_2)=\sqrt{a\Delta m^2}\times\\
    &(-i)\epsilon\int\frac{d^3p}{(2\pi)^3}\sqrt{\frac{E_{p_3}}{2}}\Big(a_p^{\scriptscriptstyle{(3)}}e^{-i\mathbf{p}\dotproduct\mathbf{x}}-a_p^{\scriptscriptstyle{(3)}\dagger }e^{i\mathbf{p}\dotproduct\mathbf{x}}\Big)\Big|_{p^0=E_{p_3}}\notag\\
    &+ig\int\frac{d^3p}{(2\pi)^3}\Bigg\{\sqrt{\frac{E_{p_1}}{2}}\Big(a_p^{\scriptscriptstyle{(1)}}e^{-i\mathbf{p}\dotproduct\mathbf{x}}-a_p^{\scriptscriptstyle{(1)}\dagger }e^{i\mathbf{p}\dotproduct\mathbf{x}}\Big)\Big|_{p^0=E_{p_1}}\notag\\
    &+\sqrt{\frac{E_{p_2}}{2}}\Big(a_p^{\scriptscriptstyle{(2)}}e^{-i\mathbf{p}\dotproduct\mathbf{x}}-a_p^{\scriptscriptstyle{(2)}\dagger }e^{i\mathbf{p}\dotproduct\mathbf{x}}\Big)\Big|_{p^0=E_{p_2}}\Bigg\}\sqrt{a\Delta m^2}.\notag
\end{align}

Now we quantize the theory canonically. For that purpose, we promote our canonical fields to operators which satisfy the equal time canonical commutation relations (\ref{relaciones_conmutacion_canonicas}) (the subindex
$3$ corresponds to $\lambda$). Employing the same procedure described in Sec.~\ref{sec:quantization-phi1-phi2}, we compute:
\begin{align}
&[a_p^{\scriptscriptstyle{(\alpha)}}, a_q^{\scriptscriptstyle{(\beta)}}]=0 \ \ , \ \ [a_p^{\scriptscriptstyle{(\alpha)}\dagger}, a_q^{\scriptscriptstyle{(\beta)}\dagger}]=0,\notag\\
&[a_p^{\scriptscriptstyle{(\alpha)}}, a_q^{\scriptscriptstyle{(\beta)}\dagger}]=0 \ \text{ si $\alpha\neq\beta$},\notag
\\
&[a_p^{\scriptscriptstyle{(2)}}, a_q^{\scriptscriptstyle{(2)}\dagger}]=-[a_p^{\scriptscriptstyle{(1)}}, a_q^{\scriptscriptstyle{(1)}\dagger}]=\frac{1}{a\Delta m^2}(2\pi)^3\delta^{(3)}(\Vec{p}-\Vec{q}),\nonumber\\
&
[a_p^{\scriptscriptstyle{(3)}}, a_q^{\scriptscriptstyle{(3)}\dagger}]=\frac{1}{\epsilon a\Delta m^2}(2\pi)^3\delta^{(3)}(\Vec{p}-\Vec{q}),
\end{align}
which are the expected commutation relations. Finally, we can obtain the Hamiltonian density via the Legendre transform $\mathcal{H}=\sigma_1\partial_0\psi_1+\sigma_2\partial_0\psi_2+\sigma_\lambda\partial_0\lambda-\mathcal{L}$. After some laborious operator algebra, we find:
\begin{align}
    \mathcal{H}&=\frac{1}{2}\left(\sigma_2^2+(\nabla(\psi_2-g\lambda))^2+m_2^2(\psi_2-g\lambda)^2)\right)\notag \\
    &-\frac{1}{2}\left(\sigma_1^2+(\nabla(\psi_1-g\lambda))^2+m_1^2(\psi_1-g\lambda)^2)\right)\\
    &+\frac{\epsilon}{2}\left(\left(\frac{1}{\epsilon}(\sigma_\lambda+g(\sigma_1+\sigma_2)\right)^2+(\nabla\lambda)^2+m_3^2\lambda^2\right),\notag
\end{align}
which yields,
\begin{align}
    H&=a\Delta m^2\int\frac{d^3p}{(2\pi)^3}\frac{E_{p_2}}{2}(a_p^{\scriptscriptstyle(2)}a_{p}^{\scriptscriptstyle(2)\dagger}+a_{p}^{\scriptscriptstyle(2)\dagger}a_{p}^{\scriptscriptstyle(2)})\notag\\
    &-a\Delta m^2\int\frac{d^3p}{(2\pi)^3}\frac{E_{p_1}}{2}(a_p^{\scriptscriptstyle(1)}a_{p}^{\scriptscriptstyle(1)\dagger}+a_{p}^{\scriptscriptstyle(1)\dagger}a_{p}^{\scriptscriptstyle(1)})\\
    &+\epsilon a\Delta m^2\int\frac{d^3p}{(2\pi)^3}\frac{E_{p_3}}{2}(a_p^{\scriptscriptstyle(3)}a_{p}^{\scriptscriptstyle(3)\dagger}+a_{p}^{\scriptscriptstyle(3)\dagger}a_{p}^{\scriptscriptstyle(3)}).\notag
\end{align}

In conclusion, the only new feature is that now fields have a new energy mode with mass $m_3$. This mass goes to infinity in the limit $\epsilon\to0^+$. Moreover, as we will see in Sec.~\ref{aparicion_phi_diagramas}, this mode is not important in the Feynman diagrams (and thus in the Green functions) because it is not present in the physical field $\phi$.

Note this Hamiltonian contains superfluous degrees of freedom: As seen in the previous Section, Lagrangian density (\ref{lagrangiano_conlambda}) contains some field combinations that could be integrated out, as they are not present in the source terms. For example, the $\lambda$ modes are completely independent and irrelevant for our original field $\phi$, which does not contain these modes.

\subsection{Fock space: ghost quantization.}
\label{estados_energia_biendefinida}
Our labor now is to construct the Fock space on which the operators present in our original Lagrangian density act. First and foremost is defining a vacuum state $\ket{0}$. From this state, the multi-particle states will be defined.

For one-particle states associated to mass $m_2$ modes we will follow the standard procedure (see, for instance, Weinberg's book~\cite{weinberg}, Sec.~1.2): we will interpret $N_p^{\scriptscriptstyle(2)}=a_p^{\scriptscriptstyle(2)\dagger}a_p^{\scriptscriptstyle(2)}$ as the number operator~\footnote{Without any loss of generality, we ignore the factor $1/(a\Delta m^2)$ in this discussion. It only affects the definition of the number operator by a global factor. We ignore, as well, the factor $(2\pi)^3\delta^{(3)}(\Vec{p}-\Vec{q})$.}
of type-2 particles of momentum $\Vec{p}$. With this and defining the action $a_p^{\scriptscriptstyle(2)}\ket{0}=0\ \forall \Vec{p}$, we obtain (using (\ref{relaciones_conmutacion_cd})) the usual result $N_2(a^{\scriptscriptstyle(2)\dagger})^n\ket{0}=n(a^{\scriptscriptstyle(2)\dagger})^n\ket{0}$. This way, we can define the one-particle state of type 2 and momentum $\Vec{p}$ as $\ket{\Vec{p}^{\scriptscriptstyle\:(2)}}=\sqrt{a\Delta m^2}\sqrt{2E_{p_2}}a_p^{\scriptscriptstyle(2)\dagger}\ket{0}$. The normalization factor is arbitrary, but we have chosen a convenient one that makes the inner product Lorentz invariant. The real positive
$\sqrt{a\Delta m^2}$ factor is added so that the $\psi_2$ field expression is in the form of the typical Klein-Gordon convention:
\begin{align}
    \Phi_\text{KG}(\mathbf{x})
    &= \int \frac{d^3p}{(2\pi)^3}\,\frac{1}{\sqrt{2E_p}}
    \Big( \text{Creator}\,e^{(\dots)}\\
        &\ \ \ \ \ \ \ \ \ \ \ \ \ \ \ \ \ \ \ \ \ \ + \text{Annihilator}\,e^{(\dots)} \Big)\bigg|_{p^0=E_p}.\notag
\end{align}
From (\ref{hamiltoniano_final}) it follows these modes contribute positively to the Hamiltonian: $H\ket{\Vec{p}^{\scriptscriptstyle\:(2)}}=(E_{p_2}+\text{V.E.})\ket{\Vec{p}^{\scriptscriptstyle\:(2)}}$.\footnote{$\text{V.E.}\equiv\text{Vacuum Energy}$.}

For states associated to the mass $m_1$ modes, which are the ghosts, the discussion is more delicate. Only two consistent ways of defining a number operator exist: {\bf (i)} $N_p^{\scriptscriptstyle(2)}=a_p^{\scriptscriptstyle(1)}a_p^{\scriptscriptstyle(1)\dagger}$, in which case $a_p^{\scriptscriptstyle(1)\dagger}$ is the annihilation operator; or {\bf (ii)} $N_p^{\scriptscriptstyle(2)}=-a_p^{\scriptscriptstyle(1)\dagger}a_p^{\scriptscriptstyle(1)}$, in which case $a_p^{\scriptscriptstyle(1)}$ is the annihilation operator. In this article, we will work with choice {\bf (i) }. However, in App.~\ref{app:ghost-number-op}, we provide a brief discussion about the two alternative options for the ghost number operator.

A similar discussion must be done for the third mode of mass $m_3$. These type-3 modes are standard modes that are not of our interest, as they are not present in $\phi(\mathbf{x})=\frac{1}{a\Delta m^2}(\psi_2(\mathbf{x})-\psi_1(\mathbf{x}))$, which is the physical field. The multi-particle states and the inner product are defined naturally as always. This way, the Fock space (which is trivially a vector space) is:
{\small\begin{align}
\mathcal{H}_\text{Fock} =& \mathop{\bigoplus}_{n,m,l=0}^\infty \Big\{
{\small\ket{\Vec{p}_1^{\scriptscriptstyle\:(1)}, \dots, \Vec{p}_n^{\scriptscriptstyle\:(1)}, \Vec{q}_1^{\scriptscriptstyle\:(2)}, \dots, \Vec{q}_m^{\scriptscriptstyle\:(2)}, \Vec{k}_1^{\scriptscriptstyle\:(3)},\dots, \Vec{k}_l^{\scriptscriptstyle\:(3)}}} \nonumber \\
& \quad  \text{s.t. } \Vec{p}_1, \dots, \Vec{p}_n, \Vec{q}_1, \dots, \Vec{q}_m, \Vec{k}_1,\dots, \Vec{k}_l \in \mathbb{R}^3 \Big\}.
\label{Fock_completo}
\end{align}}

\section{Pais-Uhlenbeck interacting Quantum Field Theory}
\label{cosas_perturbativas}

So far, we have worked with a free Pais-Uhlenbeck Quantum Field Theory. Now, we are going to explore how this theory may behave when interacting. We will assume that the appearance of ghosts does not alter the perturbation theory approach, so that the Feynman diagram formalism can still be used --essentially unchanged-- for our standard and ghost Klein-Gordon modes. That is, we are going to suppose that the perturbative expansion of the Green functions via the path integral generating functional is still valid.

In short, we will use the perturbative expansion of the Green functions, as well as the usual Feynman rules, for interactions involving $\psi_1$ and $\psi_2$, and derive some results for diagrams constructed with our Pais-Uhlenbeck field $\phi$.

\subsection{Non-derivative $\phi$ interaction}
\label{descripcion_interaccion_phi}

Let us consider the free Lagrangian density (\ref{lagrangiano}) with the additional interaction term:
\begin{eqnarray}
&&\mathcal{L}_\text{int}=\, -\,  \lambda_4 \phi^4\, .
\label{eq:phi4-int}
\end{eqnarray}
In the canonical Hawking-Hertog coordinates~(\ref{hawking-hertog}) this interaction Lagrangian reads as:
\begin{equation}
    \mathcal{L}_\text{int}=\frac{\, -\,  \lambda_4 }{[a\Delta m^2]^2}(\psi_2^4-4\psi_2^3\psi_1+6\psi_2^2\psi_1^2-4\psi_2\psi_1^3+\psi_1^4),
\end{equation}
from which we can obtain the vertices (taking into account the combinatorial factors arising from the indistinguishability of same-leg states):
\begin{equation}
\begin{aligned}
&\begin{tikzpicture}
  \begin{feynman}[scale=0.34]
    \vertex [label=left:] (a) at (-1,1);
    \vertex [label=left:] (b) at (-1,-1);
    \vertex [label=right:] (c) at (1,1);
    \vertex [label=right:] (d) at (1,-1);
    \vertex [dot] (v) at (0,0);

    \diagram* {
      (a) -- [draw] (v) -- [draw] (b),
      (c) -- [draw] (v) -- [draw] (d),
    };
  \end{feynman}
\end{tikzpicture}=-
\begin{tikzpicture}
  \begin{feynman}[scale=0.34]
    \vertex [label=left:] (a) at (-1,1);
    \vertex [label=left:] (b) at (-1,-1);
    \vertex [label=right:] (c) at (1,1);
    \vertex [label=right:] (d) at (1,-1);
    \vertex [dot] (v) at (0,0);

    \diagram* {
      (a) -- [draw] (v) -- [draw] (b),
      (c) -- [draw] (v) -- [scalar] (d),
    };
  \end{feynman}
\end{tikzpicture}=
\begin{tikzpicture}
  \begin{feynman}[scale=0.34]
    \vertex [label=left:] (a) at (-1,1);
    \vertex [label=left:] (b) at (-1,-1);
    \vertex [label=right:] (c) at (1,1);
    \vertex [label=right:] (d) at (1,-1);
    \vertex [dot] (v) at (0,0);

    \diagram* {
      (a) -- [draw] (v) -- [draw] (b),
      (c) -- [scalar] (v) -- [scalar] (d),
    };
  \end{feynman}
\end{tikzpicture}=-
\begin{tikzpicture}
  \begin{feynman}[scale=0.34]
    \vertex [label=left:] (a) at (-1,1);
    \vertex [label=left:] (b) at (-1,-1);
    \vertex [label=right:] (c) at (1,1);
    \vertex [label=right:] (d) at (1,-1);
    \vertex [dot] (v) at (0,0);

    \diagram* {
      (a) -- [draw] (v) -- [scalar] (b),
      (c) -- [scalar] (v) -- [scalar] (d),
    };
  \end{feynman}
\end{tikzpicture}=
\begin{tikzpicture}
  \begin{feynman}[scale=0.34]
    \vertex [label=left:] (a) at (-1,1);
    \vertex [label=left:] (b) at (-1,-1);
    \vertex [label=right:] (c) at (1,1);
    \vertex [label=right:] (d) at (1,-1);
    \vertex [dot] (v) at (0,0);

    \diagram* {
      (a) -- [scalar] (v) -- [scalar] (b),
      (c) -- [scalar] (v) -- [scalar] (d),
    };
  \end{feynman}
\end{tikzpicture}
%%%\notag
\\
&=
\frac{\, -\, i4!\lambda_4}{[a\Delta m^2]^2}.
\end{aligned}  \vspace*{0.25cm}
\end{equation}
Here, the continuous line denotes the standard scalar mode of mass $m_2$, and the discontinuous one the scalar ghost mode of mass $m_1$. Note the following important property: every vertex of this theory, when a standard leg is replaced by a ghost leg (or vice versa), undergoes only a sign change.

The previous discussion can be easily generalized to $\mathcal{L}_\text{int}=\, -\,  \lambda_n \phi^n$. Writing $\phi$ in terms of $\psi_1$ and $\psi_2$, using the binomial formula and taking into account the combinatorial factors that appear because of the indistinguishability of same-leg states, we conclude that the vertex of $n_1$ ghost legs and $n_2$ standard legs is:
\begin{equation}
    n!(-1)^{n_1}\frac{-\, i\lambda_n }{[a\Delta m^2]^{n/2}}\delta_{n_1+n_2 ,n}\ .
\end{equation}

Observe the factor $(-1)^{n_1}$ indicates that vertices only undergo a sign change when changing a ghost leg by a standard leg or vice-versa. Moreover, this discussion does not change if we multiply  the previous interaction by other fields of completely different nature (e.g., a photon field).

In short, if we have an arbitrary interaction where $\phi$ does not appear derived, the vertex of the theory only experiences a change of sign when changing a standard leg by a ghost leg or vice-versa. Note it is also important that the interaction does not have additional separate $\psi_1$, $\psi_2$ and $\lambda$ dependence; that is, it is key that the interaction can be completely written in terms of underived $\phi$ fields and other unrelated fields. These are the type of interactions we will have our sights on.

\subsection{The $\phi$ propagator as an internal Feynman diagram line}
\label{aparicion_phi_diagramas}

In this Subsection, we are going to study the structure of Feynman diagrams that contain  $\phi$--propagators.  First, we will discuss the case where two $\phi$--vertices are connected by this propagator. This is equivalent to examining all the possible diagrams involving the $\psi_1$, $\psi_2$ and $\lambda$ propagators that can connect these two $\phi$--vertices. In principle, one could think that there are only two possible diagrams, those with the $\psi_1$ and $\psi_2$ free un-mixed propagators. However, we must take into account the extra interaction that arises from the $\lambda$--$\psi_j$ mixing with  the Lagrange multiplier $\lambda$ in~(\ref{lagrangiano_conlambda}) (treated here as a perturbation):
\begin{equation}
\begin{aligned}
&\begin{tikzpicture}[baseline=(current bounding box.center)]
    % Línea sólida
    \draw[thick] (0,0) -- (2,0);
    % Vértice (punto negro)
    \fill (1,0) circle (2pt);
    % Etiquetas
    \node[left] at (0,0) {$\psi_2$};
    \node[right] at (2,0) {$\lambda$};

\end{tikzpicture}
&= \; i g (-p^2+m_2^2),
\\[1em]
&\begin{tikzpicture}[baseline=(current bounding box.center)]
    % Línea discontinua
    \draw[thick,dashed] (0,0) -- (1,0);
    % Vértice (punto negro)
    \fill (1,0) circle (2pt);
    % Línea sólida
    \draw[thick] (1,0) -- (2,0);
    % Etiquetas
    \node[left] at (0,0) {$\psi_1$};
    \node[right] at (2,0) {$\lambda$};

\end{tikzpicture}
&= \; - i g (-p^2+m_1^2),
\end{aligned}
\label{eq:mixing-vertices}
\end{equation}

Having this interaction in mind, the tree-level propagators that will appear between the two vertices are, then, given by the geometric sum:
\begin{equation}
\begin{aligned}
& \begin{tikzpicture}[scale=0.6]
      \begin{feynman}

        \vertex [label=left:$j$] (b) at (-2,0);

        \vertex [label=right:$j'$] (e) at (2,0);

        \vertex (v) at (0,0);

        \diagram* {

          (b) -- [draw] (v) -- [draw] (e),

        };
      \end{feynman}
      % Capa base blanca para cubrir líneas internas
  \fill[white] (0,0) circle [radius=0.5];

  % Capa con patrón de rayado
  \fill[pattern=north east lines] (0,0) circle [radius=0.5];

  % Borde sólido
  \draw[thick] (0,0) circle [radius=0.5];
    \end{tikzpicture}=\delta_{jj'}\begin{tikzpicture}=
  \begin{feynman}[scale=1]
    \vertex (a) at (0,0);  % Un vértice de inicio
    \vertex (b) at (1,0);  % Un vértice de fin
    \diagram* {
      (a) -- [draw,edge label=$j$] (b),  % Línea fotónica horizontal
    };
  \end{feynman}
\end{tikzpicture}+
\begin{tikzpicture}[baseline=-0.5ex]
  % Línea primero
  \draw (-1,0) -- (1,0);

  % extremos
  \node[left]  at (-1,0) {$j$};
  \node[right] at (1,0)  {$j'$};

  % vértices lambda
  \fill (-0.4,0) circle (1.4pt) node[xshift=12pt, yshift=-6pt] {$\lambda$};
  \fill (0.4,0) circle (1.4pt);
\end{tikzpicture}\\
&+
\begin{tikzpicture}[baseline=-0.5ex]
  % Línea primero
  \draw (-1.6,0) -- (1.6,0);

  % extremos
  \node[left]  at (-1.6,0) {$j$};
  \node[right] at (1.6,0)  {$j'$};

  % vértices lambda
  \fill (-0.9,0) circle (1.4pt) node[xshift=10pt, yshift=-6pt] {$\lambda$};
  \fill (0.9,0) circle (1.4pt) node[xshift=-10pt, yshift=-6pt] {$\lambda$};

  % círculo con Sigma que tapa la línea
  \node[draw, circle, inner sep=1.6pt, fill=white] at (0,0) {$i\Sigma$};
\end{tikzpicture}\\
&+
\begin{tikzpicture}[baseline=-0.5ex]
  % Línea base
  \draw (-2.0,0) -- (2.0,0);

  % extremos
  \node[left]  at (-2.0,0) {$j$};
  \node[right] at (2.0,0)  {$j'$};

  % vértices lambda
  \fill (-1.5,0) circle (1.4pt) node[xshift=10pt, yshift=-6pt] {$\lambda$} node[xshift=42pt, yshift=-6pt] {$\lambda$};
  \fill (1.5,0)  circle (1.4pt) node[xshift=-10pt, yshift=-6pt] {$\lambda$};

  % círculos Sigma que cubren la línea
  \node[draw, circle, inner sep=1.6pt, fill=white] at (-0.6,0) {$i\Sigma$};
  \node[draw, circle, inner sep=1.6pt, fill=white] at (0.6,0)  {$i\Sigma$};
\end{tikzpicture}+\dots \,,
\end{aligned}
\label{eq:geometric-sum}
\end{equation}
where the lines
$\begin{tikzpicture}
  \begin{feynman}[scale=1]
    \vertex (a) at (0,0);  % Un vértice de inicio
    \vertex (b) at (1,0);  % Un vértice de fin
    \diagram* {
      (a) -- [draw,edge label=$i$] (b),  % Línea fotónica horizontal
    };
  \end{feynman}
\end{tikzpicture}$
denote the un-mixed standard propagator for $i=2$ (with mass $m_2$),  and the un-mixed ghost propagator for $i=1$ (with mass $m_1$).
Likewise, we are defining:
\begin{equation}
\begin{tikzpicture}[baseline=-0.5ex]
  \node[draw, circle, inner sep=2pt, fill=white] (sigma) at (0,0) {$i\Sigma$};
\end{tikzpicture}=\begin{tikzpicture}[baseline=-0.5ex]
  % Línea primero
  \draw (-1,0) -- (1,0);

  % extremos
  \node[above]  at (-1,0) {$\lambda$};
  \node[above] at (1,0)  {$\lambda$};

  % vértices lambda
  \fill (-0.6,0) circle (1.4pt) node[xshift=17pt, yshift=6pt] {$1$};
  \fill (0.6,0) circle (1.4pt);
\end{tikzpicture}+\begin{tikzpicture}[baseline=-0.5ex]
  % Línea primero
  \draw (-1,0) -- (1,0);

  % extremos
  \node[above]  at (-1,0) {$\lambda$};
  \node[above] at (1,0)  {$\lambda$};

  % vértices lambda
  \fill (-0.6,0) circle (1.4pt) node[xshift=17pt, yshift=6pt] {$2$};
  \fill (0.6,0) circle (1.4pt);
\end{tikzpicture}\, ,
\end{equation}
being the external $\lambda$ propagators amputated
here in the definition of $\Sigma$, which takes the value $\Sigma=-g^2\Delta m^2$.  Taking into account that the un-mixed propagator of the $\lambda$ field is $\frac{i/\epsilon}{p^2-\eta}$, the geometric sum in~(\ref{eq:geometric-sum}) is found to be:
{\small\begin{equation}
\begin{tikzpicture}[scale=0.6]
      \begin{feynman}

        \vertex [label=left:$j$] (b) at (-2,0);

        \vertex [label=right:$j'$] (e) at (2,0);

        \vertex (v) at (0,0);

        \diagram* {

          (b) -- [draw] (v) -- [draw] (e),

        };
      \end{feynman}
      % Capa base blanca para cubrir líneas internas
  \fill[white] (0,0) circle [radius=0.5];

  % Capa con patrón de rayado
  \fill[pattern=north east lines] (0,0) circle [radius=0.5];

  % Borde sólido
  \draw[thick] (0,0) circle [radius=0.5];
    \end{tikzpicture}=\delta_{jj'}\frac{(-1)^ji}{p^2-m_j^2}+\frac{ig^2}{\epsilon(p^2-\eta)-g^2\Delta m^2}.
\end{equation}}

\noindent These are thus our tree-level propagators after taking into account the $\lambda$--$\psi_j$ mixing.

We are now ready to determine the structure of a Feynman diagram when $\phi$ mediates the propagation. Let
$V_A$ and $V_B$
be two vertices of our theory connected by the propagation of the $\phi$ field.
The notation
$V_{A,B}$
will refer to the vertex with respect to the internal line $\psi_2$ of the propagator, and we will have in mind the vertex property that a change $\psi_1\leftrightarrow\psi_2$ in a vertex line only changes its sign. This yields for the $\phi$--exchange diagram,
\begin{align}
    &V_A\left(\frac{i}{p^2-m_2^2}+\frac{ig^2}{\epsilon(p^2-\eta)-g^2\Delta m^2}\right) V_B\notag\\
    &+V_A\left(\frac{ig^2}{\epsilon(p^2-\eta)-g^2\Delta m^2}\right) (-V_B)\notag\\
    &+(-V_A)\left(\frac{ig^2}{\epsilon(p^2-\eta)-g^2\Delta m^2}\right) V_B\notag\\
    &+(-V_A)\left(\frac{-i}{p^2-m_1^2}+\frac{ig^2}{\epsilon(p^2-\eta)-g^2\Delta m^2}\right) (-V_B)\notag\\
    &=\, V_A\left(\frac{i}{p^2-m_2^2}+\frac{-i}{p^2-m_1^2}\right)V_B.
\end{align}
This precisely produces the $\phi$--propagator and vertices one would naively read from the original Lagrangian density~(\ref{lagrangiano}) (with $\mathcal{L}_\text{int}[\phi]$ included)
in terms of $\phi$. Indeed, the sum of the $\psi_1$ and $\psi_2$  propagators, alongside the normalization $[a\Delta m^2]^{-1}$ hidden in the vertices ($V_A V_B$), gives us the $\phi$--propagator in Eq.~(\ref{propagador}).
Notice that the remaining factors in $V_A$ and $V_B$ correspond to the aforementioned $\phi$--vertices.

If, instead of two vertices connected by a single tree level propagator, we have one vertex connected to itself through a loop, the same argument can be applied but with one vertex ($V_A$) instead of two ($V_AV_B$). Note that global symmetry factors arising from Feynman diagram overcounting do not play any role in these arguments, which focus on the structure arising from individual field contractions. This provides a strong consistency check: for a non-derivative $\phi$ interaction, the Green functions
obtained from~(\ref{FuncionalGenerador_final})
exactly coincide with those from~(\ref{generador_funcional}). In particular, the contributions from the propagators of $\psi_1$, $\psi_2$ and $\lambda$ in~(\ref{FuncionalGenerador_final}) get combined to conform the $\phi$ propagator.

\subsection{Improved UV convergence of the Green functions and renormalizability}

Let us study now the Green functions associated to our higher-derivative field beyond tree-level.
To this end, we will consider Green functions constructed solely from the $\phi$ field, which interacts with itself through a non-derivative $\phi$-interaction Lagrangian, i.e., as described in Sec.~\ref{descripcion_interaccion_phi}.
In this Subsection, we are going to analyze the improved ultraviolet convergence of these Green functions due to the faster-than-standard vanishing of the propagator at high energies, as $1/k^4$ instead of $1/k^2$.

For this purpose, we will study the Feynman diagrams that contribute to a Green function. A diagram with at least one loop will be of the form:
\begin{align}
&\int\frac{d^4k_1}{(2\pi)^4}...\frac{d^4k_L}{(2\pi)^4}\\
&\times\frac{-i/a}{(k_1^2-m_1^2)(k_1^2-m_2^2)}\ldots\frac{-i/a}{(k_P^2-m_1^2)(k_P^2-m_2^2)},\nonumber
\end{align}
where $L$ and $P$ are the number of loops and internal lines of the diagram, respectively (ignoring the external legs). Note that here we have employed that the $\phi$-interaction is non-derivative as there is no momentum dependence in the vertices. The superficial degree of divergence (defined only for loop diagrams) of this generic diagram will then be:
\begin{equation}
    D=4L-4P=4(L-P)\, .
\end{equation}

On the other hand, for any connected Feynman loop diagram, one has the relation $L=P-V+1$, with $V\geq 1$ the number of vertices of the interaction process (if $V=0$, we do not have an interaction, so we rule this case out). Using this expression in the previous result, we then obtain our master formula for the superficial degree of divergence for $V\geq1$:
\begin{equation}
    D=4(1-V)\, , \qquad \mbox{with }D\geq0 \, \iff \, V=1\, .
\end{equation}
Thus, the only superficially divergent (those with $D\geq0$) connected loop diagrams are those that can be constructed using solely one vertex. This means that only a finite number of superficially divergent loop diagrams exist. Hence, it is possible to renormalize the Green functions at all orders in perturbation theory. We will show explicitly how to do so in the following examples.

\subsubsection{Example: Renormalization of the $\phi^4$ interaction}

Let us consider again the interaction  $\mathcal{L}_\text{int}=\, -\lambda_4\phi^4$ in Eq.~(\ref{eq:phi4-int}). The only possible superficially divergent loop diagrams that can be constructed are:
\begin{equation}
\begin{tikzpicture}[scale=0.6]
  \begin{feynman}
    % Vértices
    \vertex (i) at (-2,0);     % Entrada
    \vertex (v) at (0,0);      % Vértice
    \vertex (f) at (2,0);      % Salida
    \vertex (top) at (0,1);    % Parte superior del loop

    % Diagrama
    \diagram* {
      (i) -- [photon] (v) -- [photon] (f),                                % Línea externa
      (v) -- [photon, half left, looseness=1.5] (top) -- [photon, half left, looseness=1.5] (v),  % Loop simétrico arriba
    };
  \end{feynman}
\end{tikzpicture}\ \ ,\ \
\begin{tikzpicture}[scale=0.7]
  \begin{feynman}
    % Vértice central
    \vertex (v) at (0,0);
    \vertex (left) at (-0.8,0);
    \vertex (right) at (0.8,0);

    % Diagrama
    \diagram* {
      (v) -- [photon, half left, looseness=1.5] (left) -- [photon, half left, looseness=1.5] (v),   % loop izquierdo
      (v) -- [photon, half right, looseness=1.5] (right) -- [photon, half right, looseness=1.5] (v), % loop derecho
    };
  \end{feynman}
\end{tikzpicture}\,.
\label{eq:loop-diagrams-phi4}
\end{equation}
Here we represent the $\phi$ propagator with a photon-type line. As discussed in Sec.~\ref{aparicion_phi_diagramas}, this propagator is a combination of the propagators of $\psi_1$, $\psi_2$ and $\lambda$; for this reason, we use a distinct line style to distinguish it from them.

The second diagram (vacuum bubble), although superficially divergent, does not contribute to any connected Green function of the theory (although it does contribute to the vacuum energy). Thus, regarding Green functions, the only relevant superficially diverging loop diagram is the self-energy. Any loop diagram (barring vacuum bubbles) is divergent if and only if it contains this self-energy as a sub-diagram (see \cite{peskin}, Sec. 10.1, for a more detailed discussion). Thus, the renormalization of this self-energy renormalizes the divergences of any relevant diagram of the theory, that is, the whole theory rests renormalized at all orders in the perturbative expansion.

We will adopt dimensional regularization, with space-time dimension $d=4-2\varepsilon$ (do not confuse this $\varepsilon$ with the $\epsilon$ parameter we previously introduced in the generating functional). At the end of the renormalization procedure we will take the physical limit $\varepsilon\to0$. The one-loop self-energy diagram $i\Sigma$ is given by:
\begin{eqnarray}
i\Sigma &=& - i4! \lambda_4   \frac{1 }{2}  \int\frac{d^dk}{(2\pi)^d}\widetilde{G}(\mathbf{k})
\nonumber\\
&&= \, -\, i  12 \lambda_4 \, \times \, \frac{1}{(4\pi)^2} \left(\frac{A_0(m_1^2)-A_0(m_2^2)}{a\Delta m^2}\right)
\nonumber\\
&&=\, -\,i  \frac{12\lambda_4}{(4\pi)^2 a} \left[\frac{1}{\hat{\varepsilon}}\, + C(m_j^2,\mu^2) \, \right] \,,  %%%\mbox{finite}\, ,
\label{contratermino_phi4}
\end{eqnarray}
where we have used the expression for the $\phi$--propagator in Eq.~(\ref{propagador}), taken into account the $1/2$ symmetry factor of the diagram, and finally expressed the outcome by means of the Passarino-Veltman scalar integral~\cite{Passarino:1978jh}:
\begin{eqnarray}
\frac{1}{(4\pi)^2}A_0(m^2)&\equiv& \int\frac{d^dk}{i(2\pi)^d}\frac{1}{k^2-m^2}\, , 
\end{eqnarray}
which contains a UV-divergent and a finite part, respectively given by:
\begin{eqnarray}
A_0(m^2)^{\infty} &=&  \frac{m^2}{\hat{\varepsilon}}\, ,
\\
A_0(m^2)^{\rm fin} &=& 
A_0(m^2)-A_0(m^2)^{\infty}  
\, \nonumber \\
 &\stackrel{d\to 4}{=}  &m^2\left(1- \log{\frac{m^2}{\mu^2}}\right) \, ,
\label{A0_fin}
\end{eqnarray}
with the $\overline{\rm{MS}}$ UV-divergence $\frac{1}{\hat{\varepsilon}}= \left(\frac{1}{\varepsilon}-\gamma_E+\log{4\pi}\right)\mu^{-2\varepsilon}$, being $\gamma_E\simeq 0.577$ the Euler-Mascheroni constant  and $\mu$ the arbitrary renormalization scale.
The finite constant remaining in the self-energy reads as:
\begin{align}
&C(m_j^2,\mu^2)\equiv\frac{A_0(m_1^2)^\text{fin}-A_0(m_2^2)^\text{fin}}{\Delta m^2}\\
&\stackrel{d\to 4}{=}\, 1+\frac{1}{\Delta m^2}\Bigg[m_2^2\log\left(\frac{m_2^2}{\mu^2}\right)-m_1^2\log\left(\frac{m_1^2}{\mu^2}\right)\Bigg]\, , 
\notag
\end{align}
given by the finite part of $A_0(m^2)$ in Eq.~(\ref{A0_fin}). Note that the argument $m_j^2$ refers to the $m_{1,2}^2$ dependence of the $C$ finite term.

This contribution corrects the propagator~(\ref{propagador}) in the following way:
{\small\begin{equation}
\widetilde{G}(\mathbf{p}) \,=\,     \frac{-i/a}{(\mathbf{p}^2-m_1^2)(\mathbf{p}^2-m_2^2)-\Sigma/a}\, .
\end{equation}}

To renormalize this propagator (that is, to make it finite), we need the following counterterms, which originate from the renormalization of the original bare Lagrangian:

\begin{align}
  \delta\mathcal{L}=  &-\frac{a}{2}\delta Z_\phi \phi(\Box+m_1^2+\delta m_1^2)(\Box+m_2^2+\delta m_2^2)\phi\nonumber\\
    &
      -\frac{a}{2}\delta m_1^2\phi(\Box+m_2^2)\phi
    -\frac{a}{2}\delta m_2^2\phi(\Box+m_1^2)\phi\nonumber\\
    &
     -\frac{a}{2}\delta m_1^2 \delta m_2^2 \phi^2
    -\delta\lambda_4
    Z_\phi^2\phi^4.
\label{eq:Lcounterterms}
\end{align}
This, as said, stems from the bare Lagrangian,
\begin{equation}
    \mathcal{L}=-\frac{a}{2}\phi_0(\Box+m_{1,0}^2)(\Box+m_{2,0}^2)\phi_0
    %%%+\mu^{2\epsilon}
    \, -\, \lambda_{4,0}\phi_0^4,
\end{equation}
where the bare quantities are related to the renormalized ones in the form:
\begin{eqnarray}
&&  \phi_0=Z_\phi^{1/2}\phi=(1+\delta Z_\phi)^{1/2}\phi\,, \qquad m_{i,0}^2=m_i^2+\delta m_i^2\,,
    \nonumber\\
&&\lambda_{4,0}=
\lambda_4+\delta\lambda_4\, .
\end{eqnarray}

The self-energy renormalization requires the cancellation of the UV-divergences when the counterterm contributions~(\ref{eq:Lcounterterms}) are summed up to the divergent loop diagram $i\Sigma$:
\begin{eqnarray}
i\Sigma^r&=&-i a \, \delta  Z_\phi   (p^2-m_1^2-\delta m_1^2)(p^2-m_2^2-\delta m_2^2) \nonumber\\
    &&
     +ia\, \delta m_1^2\,  (p^2-m_2^2)
    +ia\, \delta m_2^2 \, (p^2-m_1^2)\nonumber\\
    &&
    -ia\delta m_1^2 \delta m_2^2 \, +\,   i\Sigma \nonumber\\
    &&
= \mbox{finite}\, .
\label{eq:SE-renormalization}
\end{eqnarray}
Since $i\Sigma$ is $\mathbf{p}$-independent --this is, $p^4$ and $p^2$ divergent terms are absent in~(\ref{contratermino_phi4}) -- our renormalization conditions must obey (in the $\overline{\rm{MS}}$-scheme):
\begin{eqnarray}
    \delta Z_\phi=0\,,\qquad \delta m_1^2=-\delta m_2^2\equiv\delta M^2\, .
\end{eqnarray}
Finally, the self-energy renormalization is achieved by imposing that the constant $\mathcal{O}(p^0)$ divergent term also vanishes, leading to the condition:
\begin{equation}
    (\delta M^2)^2+\Delta m^2\delta M^2 \, =\,\frac{3\lambda_4}{4\pi^2 a^2}  
    \frac{1}{\hat{\varepsilon}}\,,
\end{equation}
in the $\overline{\rm{MS}}$--scheme. Notice that $\delta\lambda_4$ is UV-finite since the 4-point Green function and the wave-function renormalization $Z_\phi$ are finite. Hence, one has $Z_\phi=1$ and $\lambda_{4,0}=\lambda_4$ in the $\overline{\rm{MS}}$--scheme.

\subsubsection{Example: Renormalization of the $\phi^6$ interaction}

Let us now consider the non-derivative interaction:
\begin{equation}
    \mathcal{L}_\text{int}\, = \, -\,  \lambda_4\phi^4 \, -\, \lambda_6\phi^6\,.
\end{equation}

The relevant superficially divergent loop diagrams are:
\begin{equation}
\begin{tikzpicture}[scale=0.6]
  \begin{feynman}
    % Vértices
    \vertex (i) at (-2,0);     % Entrada
    \vertex (v) at (0,0);      % Vértice
    \vertex (f) at (2,0);      % Salida
    \vertex (top) at (0,1);    % Parte superior del loop

    % Diagrama
    \diagram* {
      (i) -- [photon] (v) -- [photon] (f),                                % Línea externa
      (v) -- [photon, half left, looseness=1.5] (top) -- [photon, half left, looseness=1.5] (v),  % Loop simétrico arriba
    };
  \end{feynman}
\end{tikzpicture}\ \ ,\ \
\begin{tikzpicture}[scale=0.43]
  \begin{feynman}
    % Vértice central
    \vertex (v) at (0,0);

    % Vértices externos
    \vertex (a) at (-2,0);
    \vertex (b) at (2,0);

    % Vértices para los loops
    \vertex (u1) at (-0.7,0.7);
    \vertex (u2) at (0.7,0.7);
    \vertex (d1) at (-0.7,-0.7);
    \vertex (d2) at (0.7,-0.7);

    % Diagrama
    \diagram* {
      (a) -- [photon] (v) -- [photon] (b),
      (v) -- [photon] (u1),
      (v) -- [photon] (u2),
      (u1) -- [photon, half left, looseness=1.5] (u2),
      (v) -- [photon] (d1),
      (v) -- [photon] (d2),
      (d1) -- [photon, half right, looseness=1.5] (d2),
    };
  \end{feynman}
\end{tikzpicture}\ \ ,\ \
\begin{tikzpicture}[scale=0.45]
  \begin{feynman}
    % Vértice central
    \vertex (v) at (0,0);

    % Vértices externos
    \vertex (a) at (-1,-1);
    \vertex (b) at (1,-1);
    \vertex (c) at (-1,0);
    \vertex (d) at (1,0);

    % Vértices para el loop en la parte superior
    \vertex (u1) at (-0.5,1);
    \vertex (u2) at (0.5,1);

    % Diagrama
    \diagram* {
      (v) -- [photon] (a),
      (v) -- [photon] (b),
      (v) -- [photon] (c),
      (v) -- [photon] (d),
      (v) -- [photon] (u1),
      (v) -- [photon] (u2),
      (u1) -- [photon, half left, looseness=1.5] (u2),
    };
  \end{feynman}
\end{tikzpicture}\,.
\label{eq:phi6-loops}
\end{equation}
There are two additional superficially divergent loop diagrams, but they are vacuum bubbles. Again, any relevant loop diagram of the theory diverges if and only if it contains one of the previous diagrams as a sub-diagram (their renormalization implies, thus, renormalizing the whole theory at all orders). The first and second diagrams contribute to the self-energy, and they will be both addressed by the mass renormalization, as before. Regarding the third one, in order to cancel its UV-divergence we must renormalize the $\lambda_4$ operator, being its inclusion necessary to render the theory finite.

The first diagram in~(\ref{eq:phi6-loops}) gives the contribution $i\Sigma^{(2)}_1$, with the structure provided by Eq.~(\ref{contratermino_phi4}).
However, note that the $\delta\lambda_4$ counterterm also generates a superficially divergent loop diagram at higher orders, of the same kind as the one produced by the renormalized interaction $\lambda_4$. That contribution must also be
taken into account in the all-order renormalization procedure, as we also need to cancel out the divergence of the following diagram: 
\begin{equation}
\begin{tikzpicture}[scale=0.7]
  \begin{feynman}
    % Vértices
    \vertex (i) at (-2,0);     % Entrada
    \vertex (v) at (0,0);      % Vértice
    \vertex (f) at (2,0);      % Salida
    \vertex (top) at (0,1);    % Parte superior del loop

    % Diagrama
    \diagram* {
      (i) -- [photon] (v) -- [photon] (f),
      (v) -- [photon, half left, looseness=1.5] (top) -- [photon, half left, looseness=1.5] (v),
    };
  \end{feynman}

  % Círculo blanco con cruz (contratérmino)
  \draw[fill=white, thin] (0,0) circle (0.15);
  \draw[thin] (-0.1,-0.1) -- (0.1,0.1);
  \draw[thin] (-0.1,0.1) -- (0.1,-0.1);
\end{tikzpicture}  \,,
\label{eq:loop-diagrams-phi4_ct}
\end{equation}
which is the only superficially  divergent loop diagram that arises from the counterterms.

The second and third diagrams in~(\ref{eq:phi6-loops}), respectively $i\Sigma_2^{(2)}$ and $i\Sigma^{(4)}$, are derived in an analogous manner, being provided by the expressions:
\begin{eqnarray}
i\Sigma^{(2)}_2 &=& \, -\, i6!\lambda_6\frac{1}{8} \left(\int\frac{d^dk}{(2\pi)^d}\widetilde{G}(\mathbf{k})\right)^2
\nonumber\\
&&= \, -\, i  90 \lambda_6 \, \times \, \left[\frac{1}{(4\pi)^2} \left(\frac{A_0(m_1^2)-A_0(m_2^2)}{a\Delta m^2}\right)\right]^2
\nonumber\\
&&=\, -\,i  \frac{90\lambda_6}{(4\pi)^4 a^2} \left[ \frac{1}{\hat{\varepsilon}}\, +\, 
    C(m_j^2,\mu^2) \right]^2\, ,
\label{contratermino_phi6_2}
\\
i\Sigma^{(4)}&=&
\, -\, i6!\lambda_6\frac{1}{2} \int\frac{d^dk}{(2\pi)^d}\widetilde{G}(\mathbf{k})
\nonumber\\
&&= \, -\, i360 \lambda_6 \, \times \,  \frac{1}{(4\pi)^2} \left(\frac{A_0(m_1^2)-A_0(m_2^2)}{a\Delta m^2}\right)
\nonumber\\
&&=\, -\,i  \frac{360\lambda_6}{(4\pi)^2 a}  \left[ \frac{1}{\hat{\varepsilon}}\, +\, 
    C(m_j^2,\mu^2) \right] \, .
\label{contratermino_phi6_4}
\end{eqnarray}
Note that these two diagrams have a different symmetry factor: $1/8$ for $i\Sigma_2^{(2)}$ and $1/2$ for $i\Sigma^{(4)}$.

For the renormalization of the self-energy, we follow the same procedure as before:
we demand that the counterterms cancel out the UV-divergences of the loop diagrams $\left(i\Sigma|_{\lambda_4}+i\Sigma|_{\delta\lambda_4}+i\Sigma_2^{(2)}\right)$ (respectively, first diagram in~(\ref{eq:phi6-loops}), diagram~(\ref{eq:loop-diagrams-phi4_ct}) and second diagram in~(\ref{eq:phi6-loops})).
This gives us the renormalization condition~(\ref{eq:SE-renormalization}), but now with the combination $\left(i\Sigma|_{\lambda_4}+i\Sigma|_{\delta\lambda_4}+i\Sigma_2^{(2)}\right)$ instead of simply $i\Sigma$.
However, since that combination still lacks $p^4$ and $p^2$ divergent terms, the masses and wave-function renormalizations must also obey:
\begin{eqnarray}
    \delta Z_\phi=0\,,\qquad \delta m_1^2=-\delta m_2^2\equiv\delta M^2\, ,
\end{eqnarray}
in the $\overline{\rm{MS}}$-scheme.
Requiring that the constant $\mathcal{O}(p^0)$ divergent term vanishes yields the condition: 
\begin{align}
    &(\delta M^2)^2+\Delta m^2\delta M^2 \, =\, \frac{3\lambda_4}{4\pi^2 a^2}
    \frac{1}{\hat{\varepsilon}}\label{eq: mass_ren}\\
&+\frac{3\delta\lambda_4}{4\pi^2 a^2}\left[ 
    \frac{1}{\hat{\varepsilon}}+
    C(m_j^2,\mu^2)\right]\notag\\
    &\,+ \,
     \frac{90\lambda_6}{(4\pi)^4 a^{3}
     }\left[ 
    \frac{1}{\hat{\varepsilon}^2}+\frac{2C(m_j^2,\mu^2)}{\hat{\varepsilon}}\right] 
    \,.\notag
\end{align}
Note that the $C(m_j^2,\mu^2)$ constant is non-local with respect to the massive parameters of the theory. The crossed term of the square in~(\ref{contratermino_phi6_2}) gives place to a non-local sub-divergence of the form $C(m_j^2,\mu^2)/\hat{\varepsilon}$ that needs to be cancelled out with a one-loop countertem diagram. We will see that the $\delta\lambda_4$ counterterm inserted in the one-loop diagram~(\ref{eq:loop-diagrams-phi4_ct}) yields a contribution of the form $C(m_j^2,\mu^2)\delta\lambda_4\sim C(m_j^2,\mu^2)/\hat{\varepsilon}$, since $\delta\lambda_4\sim1/\hat{\varepsilon}$. Therefore, the explicit finite part of the one-loop function must be kept and combined with the divergent counterterm in order to get the cancellation of the non-local sub-divergence.

In the present theory, the $i\Sigma^{(4)}$ loop diagram contributing to the 4-point Green function is also UV-divergent. Thus, its cancellation requires a non-zero $\delta\lambda_4$ counterterm given by:
\begin{equation}
    \delta\lambda_4=-\frac{15\lambda_6}{(4\pi)^2a}\frac{1}{\hat{\varepsilon}}\,,
    \label{eq: deltalambda4}
\end{equation}
in the $\overline{\rm{MS}}$-scheme, where we took into account that $Z_\phi=1$. Notice that this time $\delta\lambda_6$ is UV-finite since the 6-point Green function and the wave-function renormalization $Z_\phi$ are finite, and, therefore, one has $Z_\phi=1$ and $\lambda_{6,0}=\lambda_6$ in the $\overline{\rm{MS}}$-scheme.

Having established the form of $\delta\lambda_4$, we now return to the discussion of mass renormalization. Substituting Eq.~(\ref{eq: deltalambda4}) into~(\ref{eq: mass_ren}) we obtain:
\begin{eqnarray}
    (\delta M^2)^2+\Delta m^2\delta M^2 \, =\, \frac{3\lambda_4}{4\pi^2 a^2}
    \frac{1}{\hat{\varepsilon}}-\frac{90\lambda_6}{(4\pi)^4 a^3
    }\frac{1}{\hat{\varepsilon}^2} \,,
\end{eqnarray}
in the $\overline{\rm{MS}}$-scheme. 
Observe the non-local divergences cancel each other out, as it should be. Indeed, these types of divergences must disappear for us to have a consistently renormalized theory (see Weinberg's non-locality theorem \cite{Weinberg1960, HahnZimmermann1968} in, e.g., Ref.~\cite{collins}, Sec.~5.8).

\subsubsection{Renormalization of the $\phi^{n}$ interaction}

To renormalize $\phi^{2n}$ with $n\in\mathbb{N}_+$ we must insert
all the non-derivative $\phi$ operators with lower even powers: $\phi^{2n-2},\phi^{2n-4},...,\phi^4$ and $\phi^2$, the last one being already included in (\ref{lagrangiano}) from the beginning by virtue of the mass terms. In other words, the following interaction is renormalizable:
\begin{equation}
\mathcal{L}_\text{int}\, =\, -\, \sum\limits_{k=1}^n \lambda_{2k}\phi^{2k}.
\end{equation}
A similar argument holds for $\phi^{2n+1}$, but considering these types of interactions, which yield unbounded potentials, arises new problems we have not treated in this work (tadpoles, spontaneous symmetry breaking, etc).

In conclusion, we observed that the improved UV-convergence of the higher-derivative $\phi$-propagator leads to a wide class of unexpectedly renormalizable theories: QFT's with only $\phi$ fields and non-derivative interactions have renormalizable Green functions (in the sense that only a finite number of operators must be introduced to eliminate the divergences at all orders in perturbation theory). This occurs even in theories that would not be renormalizable in the usual case (that is, with a standard kinetic term) like, e.g., $\phi^6$.

\section{Conclusions}
\label{sec:conclusions}

In this work, we have carried out a thorough study of the Pais–Uhlenbeck oscillator \cite{Pais:1950za, Pavsic:2016ykq,Pavsic:2013noa, Smilga:2008pr} in its QFT formulation~\cite{Gibbons:2019lmj}. First of all, we have introduced the classical, non-relativistic version of the system. Using Ostrogradsky canonical coordinates~\cite{Chen:2012au}, we have obtained the corresponding Hamiltonian, finding that it is unbounded. This serves as a clear example of Ostrogradsky's instability~\cite{Ostrogradsky:1850fid}, present in all classical higher-derivative theories.

Once understood the problematic, we started from the QFT version of the Pais-Uhlenbeck oscillator. Using Ostrogradsky canonical coordinates, we have obtained the central dynamical objects of the theory:  the EoM, the propagator, the commutation relations and the Hamiltonian.
We then perform a canonical transformation to the more convenient Hawking–Hertog canonical coordinates~\cite{Hawking:2001yt}. In addition to verifying our previous findings, the great advantage of these $(\psi_1,\psi_2)$ coordinates is that they essentially separate the ghost and standard-mode creation and annihilation operators in the quantization process. In comparison, Ostrogradsky's coordinates $(\phi_1,\phi_2)$ widely mix these two modes.

Subsequently, we constructed the path-integral generating functional of the theory. We observed, by employing functional identities, that the inclusion of canonical coordinates introduces an additional term in the Lagrangian density. This extra term corresponds to an auxiliary Lagrange multiplier field $\lambda(\mathbf{x})$.
We have checked that this term does not give rise to
issues in the canonical quantization treatment. All the derivation was performed in Hawking-Hertog coordinates, where we proved that, ultimately, the Lagrange multiplier has no impact on the total Hamiltonian of the theory, as expected. In summary, the computation of the Green functions from the properly defined path integral generating functional is found to be fully identical to what one has in standard QFT's. Finally, we have discussed the construction of the Fock space of the free theory.

To conclude, we have shown the benefits of having a higher-derivative kinetic term in an interacting theory. It has been shown that the aforementioned Lagrange multiplier $\lambda(\mathbf{x})$ plays no role beyond the consistent quantization of the canonical coordinates: the Green functions are ultimately independent of the $\lambda(\mathbf{x})$ normalization $\epsilon$, mass $\eta$ and mixing $g$ in the generating functional~(\ref{FuncionalGenerador_final}). The relevant combination of propagators in canonical coordinates exactly reproduces the $\phi$ propagator in the higher-derivative Lagrangian,
as expected.

We have also shown that  the improved UV-convergence of the loop integrals renders theories with generic non-derivative $\phi^{2n}$ interactions renormalizable for any $n\in \mathbb{N}_+$.
Contrary to what happens in theories with standard kinetic terms, we observe that the addition of non-derivative $\phi$ operators $\mathcal{O}_\phi$ of canonical dimensions $D_{\mathcal{O}_\phi}>4$ does not spoil renormalizability.

In summary, we have shown that the path integral and Green function construction is fully consistent and well defined. Nonetheless, we are perfectly aware of the several important problems that still remain open, such as, e.g., the unbounded energy spectrum.
In a future work, we will build upon this groundwork by addressing aspects that were not explored in depth here, such as potential restrictions on the Fock space in the spirit of Gupta-Bleuler~\cite{Gupta:1949rh,Bleuler:1950cy,greiner} and a more detailed investigation of the perturbative treatment.

\section*{Acknowledgments} This work is partially supported by the projects PID2022-139841NB-I00 and PID2022-137003NB-I00 funded by MICIU/AEI/10.13039/501100011033 and by ERDF/EU. This work is also part of the COST (European Cooperation in Science and Technology) Actions CA21106, CA21136, CA22113 and CA23130.

\appendix

\section{Additional functional identities}
\label{app:further-functional-details}

In order to simplify the form of the generating functional (\ref{generador_funcional}) after inserting (\ref{eq:identity-deltas}) in the measure, it has been necessary to temporarily invert the operators $(\Box+m_i^2)$ to employ Eq. (\ref{eq:delta-relation}). More explicitly:
{\small\begin{align}
  &\int D\phi \;
    \delta\!\left(
      \psi_{1}-\sqrt{\tfrac{a}{\Delta m^{2}}}\,(\Box+m_{2}^{2})\phi
    \right)
    \delta\!\left(
      \psi_{2}-\sqrt{\tfrac{a}{\Delta m^{2}}}\,(\Box+m_{1}^{2})\phi
    \right)\notag \\
  &= \frac{\Delta m^{2}}{a}
     (\Box+m_{1}^{2})^{-1}(\Box+m_{2}^{2})^{-1}
     \int D\phi \; \times\notag \\
  &\delta\!\left(
      \phi-\sqrt{\tfrac{\Delta m^{2}}{a}}
      (\Box+m_{2}^{2})^{-1}\psi_{1}
    \right)
    \delta\!\left(
      \phi-\sqrt{\tfrac{\Delta m^{2}}{a}}
      (\Box+m_{1}^{2})^{-1}\psi_{2}
    \right)\notag \\
  &= \frac{\Delta m^{2}}{a}
     (\Box+m_{1}^{2})^{-1}(\Box+m_{2}^{2})^{-1}
     \delta\!\Biggl(
       \sqrt{\tfrac{\Delta m^{2}}{a}}
       (\Box+m_{1}^{2})^{-1}\psi_{2}\notag
       \\
       &-\sqrt{\tfrac{\Delta m^{2}}{a}}
       (\Box+m_{2}^{2})^{-1}\psi_{1}
     \Biggr)\\
     &= \sqrt{\frac{\Delta m^2}{a}}\delta\!\left(
      (\Box+m_{2}^{2})\psi_{2}
      -(\Box+m_{1}^{2})\psi_{1}
    \right).\notag
\\
&
  = \sqrt{\frac{\Delta m^2}{a}}
\int D\lambda \, e^{ i \lambda \left(
      (\Box+m_{2}^{2})\psi_{2}
      -(\Box+m_{1}^{2})\psi_{1}
    \right) }\, .
\notag
\end{align}}

\noindent After this, the delta function was finally written in the last line in a path integral form. The proper normalization of the delta is hidden in the integral measure and will not be discussed any further.

\section{ Integration of the Lagrange multiplier and recovery of the original Lagrangian }
\label{app:apendice1}

We will depart from Eq.~(\ref{funcional_generador_3campos}). We then perform the change of variables from $(\psi_1,\psi_2,\lambda)$ to $(\varphi_1,\varphi_2,\lambda)$, where $\varphi_{1,2}=\frac{1}{\sqrt{2}}(\psi_2\mp\psi_1)$. This transformation has a unit Jacobian, so no additional constants appear in the measure. Performing this change of variables in the integrand yields:
\begin{align}
    &\int D\psi_1D\psi_2D\lambda\; e^{
    i\int d^4\boldsymbol{x} \Big(\frac{1}{2}\frac{\Delta m^2}{2}(\varphi_1^2+\varphi_2^2)-\varphi_1\left(\Box+\frac{m_1^2+m_2^2}{2}\right)\varphi_2
    \Big)
    }  \nonumber \\
    &\times e^{ i\int d^4\boldsymbol{x}\sqrt{2}\lambda\left(\Box\varphi_1+\frac{m_1^2+m_2^2}{2}\varphi_1-\frac{\Delta m^2}{2}\varphi_2\right)+
    i\int d^4\boldsymbol{x}
         J\sqrt{\frac{2}{a\Delta m^2}}\varphi_1
    }.
\end{align}
Now we can complete the square in the action to integrate out $\varphi_2$. The translation (with an obviously unit Jacobian) that makes this possible is:
\begin{equation}
    \varphi_2'=\varphi_2-\frac{\left(\Box+\frac{m_1^2+m_2^2}{2}\right)\varphi_1+\sqrt{2}\frac{\Delta m^2}{2}\lambda}{2\frac{1}{2}\frac{\Delta m^2}{2}}.
\end{equation}
This results in:
\begin{align}
    &\int D\varphi_2' e^{i\int d^4\mathbf{x}\frac{1}{2}\frac{\Delta m^2}{2}(\varphi_2')^2}
    \label{eq:decoupled-gen-funct}
    \\
    &\times\int D\varphi_1 D\lambda
    e^{
    i\int d^4\boldsymbol{x} \Big(\frac{2}{\Delta m^2}\frac{-1}{2}\varphi_1(\Box+m_1^2)(\Box+m_2^2)\varphi_1
    \Big)
    }  \notag \\
    &\times e^{ i\int d^4\boldsymbol{x}\frac{-\Delta m^2}{2}\lambda^2+
    i\int d^4\boldsymbol{x}
         J\sqrt{\frac{2}{a\Delta m^2}}\varphi_1\notag
    }.
\end{align}

Note the $\lambda$ field can be now trivially integrated out as well. Performing another change of variables $\phi=\sqrt{\frac{2}{a\Delta m^2}}\varphi_1$ we arrive at:
{\small\begin{align}
    &\int D\varphi_2' e^{i\int d^4\mathbf{x}\frac{1}{2}\frac{\Delta m^2}{2}(\varphi_2')^2}\int D\lambda e^{ i\int d^4\boldsymbol{x}\frac{-\Delta m^2}{2}\lambda^2}\notag\\
    &\times\int D\phi
    e^{
    i\int d^4\boldsymbol{x} \frac{-a}{2}\phi(\Box+m_1^2)(\Box+m_2^2)\phi+i\int d^4\boldsymbol{x}
         J\phi}.
\label{eq:gen-funct-from-multiplier}
\end{align}}Here we have ignored global irrelevant Jacobian constants. %If we wished to compute the integrals of the first line in every time and momentum slice, we could do so using Fresnel integral formula.\footnote{$\int_{-\infty}^\infty e^{\pm iax^2}dx=\sqrt{\frac{\pi}{a}}e^{\pm i\pi/4}\ \ \forall a\in\mathbb{R}_+$}.
Based on the relation $a\Delta m^2>0$, which means that $a$ and $\Delta m^2$ share the same sign, we observe that the $\lambda$ field contributes to the Euclidean action with the same sign as the $\phi$ term. On the other hand, $\varphi'_2$ always contributes with the opposite sign.
The $\lambda$ field does not provoke the boundlessness of the Euclidean action: it is the $\varphi_2'$ field that does.
Fortunately, both fields are decoupled from both the source $J$ and any field. Thus, their integration only contributes with a constant factor to the global (irrelevant) normalization of the generating functional,   leaving $\phi$ as the only relevant dynamical field of the theory.

It is interesting to note that while $\lambda$ has a non-zero (although trivial) contribution to the system in canonical quantization when $\epsilon>0$, this is not the case for $\varphi_2'$. Indeed, this field is:
\begin{equation}
    \varphi_2'=\frac{1}{\sqrt{2}}\frac{2}{\Delta m^2}\Big[(\Box+m_1^2)\psi_1-(\Box+m_2^2)\psi_2\Big]-\sqrt{2}\lambda,
\end{equation}
which evaluates to zero using the kinematic equations.

\section{On the choice of the ghost number operator}
\label{app:ghost-number-op}
In this Appendix, we want to briefly discuss the two alternative options one finds in the procedure of defining the ghost number operator. For example,
if we took option {\bf (i)} referred to in the previous Sec. \ref{estados_energia_biendefinida},
we would define, then, $a_p^{\scriptscriptstyle(1)\dagger}\ket{0}=0\ \forall \Vec{p}$ and $\ket{\Vec{p}^{\scriptscriptstyle\:(1)}}=-\sqrt{a\Delta m^2}\sqrt{2E_{p_1}}a_p^{\scriptscriptstyle(1)}\ket{0}$. We similarly would find $H\ket{\Vec{p}^{\scriptscriptstyle\:(1)}}=(-E_{p_1}+\text{V.E.})\ket{\Vec{p}^{\scriptscriptstyle\:(1)}}$.
In this case, these ghost states would contribute with negative energy.

The other way of defining ghost states would be option {\bf (ii)}, that is, taking the operator $a_p^{\scriptscriptstyle(1)}$ as the annihilator and defining $\ket{\Vec{p}^{\scriptscriptstyle\:(1)}}=-\sqrt{a\Delta m^2}\sqrt{2E_{p_1}}a_p^{\scriptscriptstyle(1)\dagger}\ket{0}$. This results into type-1 states of positive energy, but negative squared norm~\cite{weinberg}. Although this conception of ghosts may seem unnatural at first, it is used by many authors \cite{Salvio:2015gsi, Mannheim:2006rd}.

Each option has its advantages and disadvantages. For option {\bf (i)} (negative energy, positive squared norm), the Fock space is at least a pre-Hilbert space. Indeed, the inner product would be a sesquilinear form satisfying $\braket{\chi|\psi}=\overline{\braket{\psi|\chi}}$, $\braket{\psi|\psi}=0\iff\ket{\psi}=0$ and semi-definite positiveness. The last property is only possible because squared norms have been chosen to be non-negative. For option {\bf (ii)} (positive energy, negative squared norm), one renounces to the pre-Hilbert space construction (thus losing mathematical properties and compromising the probabilistic interpretation of Quantum Mechanics) of the Fock space, as negative squared norms are allowed. However, for this price one gains a great physical advantage: the Hamiltonian spectrum becomes bounded from below.

Other approaches to the previous two realizations of ghost states exist. However, they require non-standard techniques. For example, Refs.~\cite{Bender:2007wu, Bender:2008vh} consider modifications of the inner product.

One could argue that the anomalous behavior of the ghost modes could be improved if one restricted the Fock space in a similar way Gupta-Bleuler does in Quantum Electrodynamics~\cite{Gupta:1949rh,Bleuler:1950cy} (see, for instance, Ref.~\cite{greiner}, Sec.~7.4). This would mean that only some combination of the energy modes in the Fock space might be truly physical. However, this could give rise to unitarity problems, also frequent in ghost-theories. This issue is beyond the scope of this article and is left for a future work.

%%%%%%%%%%%%%%%%%%%
%% BIBLIOGRAFÍA
%%%%%%%%%%%%%%%%%%%

\end{document}